\documentclass[aps,twocolumn,superscriptaddress,longbibliography,floatfix,notitlepage]{revtex4-1}

\usepackage[utf8x]{inputenc}
\usepackage[english]{babel}
\usepackage[T1]{fontenc}
\usepackage{lmodern}

\usepackage{dsfont}
\usepackage{amsmath,amsfonts,amssymb,amsthm,bm,times}
\usepackage[colorlinks={true}, citecolor={blue}, filecolor={blue}, linkcolor={blue}, urlcolor={blue}]{hyperref}
\usepackage{graphicx,color}
\usepackage[caption=false]{subfig}
\usepackage{microtype}
\usepackage{braket}

\usepackage{dcolumn}
\usepackage{bm}

\begin{document}

  \title{Heat Bath Algorithmic Cooled Quantum Otto Engines}

  \author{Emre K\"{o}se}
  \affiliation{Department of Physics, Ko\c{c} University, 34450 Sariyer, İstanbul, Turkey}

  \author{Sel\c{c}uk \c{C}akmak}
  \affiliation{Department of Physics, Ko\c{c} University, 34450 Sariyer, İstanbul, Turkey}
  
  \author{Azmi Gen\c{c}ten}
  \affiliation{Department of Physics, Ondokuz Mayıs University, 55139 Samsun, Turkey}

  \author{Iannis K. Kominis}
  \affiliation{Department of Physics, University of Crete, 70013 Heraklion, Greece}

  \author{\"{O}zg\"{u}r E. M\"{u}stecapl{\i}o\u{g}lu}
  \email{omustecap@ku.edu.tr}
  \affiliation{Department of Physics, Ko\c{c} University, 34450 Sariyer, İstanbul, Turkey}

    \date{\today}
\begin{abstract}
  We suggest alternative quantum Otto engines, using heat bath algorithmic cooling with partner pairing algorithm instead of isochoric cooling. Liquid state nuclear magnetic resonance systems in one entropy sink are considered as working fluids. Then, the extractable work and thermal efficiency are analyzed in detail for four-stroke and two-stroke type of quantum Otto engines.  The role of heat bath algorithmic cooling in these cycles is to use a single entropy sink instead of two. Also, this cooling algorithm increases the power of engines reducing the time required for one cycle.
\end{abstract}
\newpage
\maketitle

\section{Introduction}
With the advances of miniaturized information and energy devices, the question of whether using
a quantum system to harvest a classical resource can have an advantage over a classical harvester
has gained much attention in recent years~\cite{PhysRevE.76.031105, Huang2014, Thomas2014,PhysRevA.75.062102,1402-4896-88-6-065008,PhysRevE.83.031135,Scully,Zhang2008,
PhysRevE.83.031135,Cakmak2016,PhysRevLett.112.150602,
PhysRevE.93.012145,0295-5075-118-6-60003,PhysRevE.95.032111,PhysRevE.76.031105,PhysRevX.5.031044,Alecce2015,PhysRevLett.2.262,0305-4470-12-5-007,PhysRevLett.105.130401,1367-2630-16-9-095003,PhysRevE.87.012140,doi:10.1146/annurev-physchem-040513-103724, PhysRevE.68.016101, Ronagel325,PhysRevLett.109.203006}. 
The argument is more or less settled in the case of quantum information devices,
and the main challenge remained is their efficient implementation. In typical quantum information devices, both the inputs and the algorithmic steps of operation are of completely quantum nature. In contrast, quantum energy devices process incoherent inputs and they operate with thermodynamical processes being quantum analogs of their classical counterparts. The quantum superiority in such a thermal device reveals itself when the resource has some quantum character, for example squeezing~\cite{PhysRevX.7.031044,PhysRevLett.112.030602}, or when the harvester has profound
quantum nature, for example, quantum correlations~\cite{0295-5075-88-5-50003,Dag2016}. Studies of both cases are limited to machine processes analogs of classical thermodynamical ones. Here we ask how can we use genuine quantum steps in the machine operation and if we can do so what are the quantum advantages we can get. 

As a specific system to explore completely quantum steps in thermal quantum device operation we consider an NMR
system. Very recently NMR quantum heat engines become experimentally available~\cite{Peterson2018,DeAssis2018}. Power outputs of these machines are not optimized. One can use non-classical resources, though such a resource would not be natural and require
some generation cost reducing overall efficiency. Alternatively one can use dynamical shortcuts to speeding up adiabatic transformations~\cite{Cakmak2017} but this would increase experimental complexity, and moreover in NMR thermalization is more seriously slow step reducing the power output of the NMR machines.

Here we, propose to replace adiabatic steps by SWAP operations while the cooling step by an algorithmic cooling~\cite{Park2016,PhysRevLett.116.170501,PhysRevA.93.012325,Brassard20144,Brassard2014,Fernandez2004,Elias,Boykin3388,Kafri2012,PhysRevLett.114.100404,Rodriguez-Briones2017b,PhysRevLett.100.140501,PhysRevLett.94.120501,Elias2006}. By this way, NMR thermal device operation would closely resemble an NMR quantum computer~\cite{OLIVEIRA2007137,Devoret2011,Abragam1963,1751-8121-49-14-143001} albeit processing a completely noisy input. We remark that
there will be only the classical energy source as the input to the machine, while the second heat bath required by the second law of thermodynamics for work production would be an effective one, engineered by a spin ensemble, typical in the algorithmic cooling scheme. In addition to engine cycles, NMR systems were also proposed for studies of single-shot thermodynamics~\cite{PhysRevLett.113.140601}. 

Our scheme allows us to provide at least one answer, in the context of NMR heat engines, to the question of how to implement genuine quantum steps in
quantum machine operation to harvest a classical energy source. In addition, our calculations suggest that compared to the NMR engine with standard thermodynamical steps quantum algorithmic NMR engine produce more power.
The advantage comes from the replacing the long-time isochoric cooling process with the more efficient
and fast algorithmic cooling stage. We provide systematic investigation by first examining the case of standard Otto cycle as a benchmark then introduce the algorithmic cooling stage instead of isochoric cooling one, and finally introduce the SWAP operation stages instead of the adiabatic transformations.

The organization of the paper is as follows. We review the theory and heat bath algorithmic cooling (HBAC) considering 3-qubit NMR sample in Sec.~\ref{sec2}. The results and discussions are given in Sec.~\ref{sec3}. In Sec.~\ref{sec3a}, efficiency, work, and power output of four-stroke quantum Otto cycles cooled by HBAC and isochoric stage are discussed by considering the same parameters to extract work. Two-stroke type engine results are discussed in Sec.~\ref{sec3b}. We conclude in Sec.~\ref{con}. The details of the HBAC using partner pairing algorithm (PPA) are given in Appendix \ref{apx}. 
\section{THE WORKING FLUID}
\label{sec2}
\begin{figure}[h!]
  \includegraphics[width=0.7\linewidth]{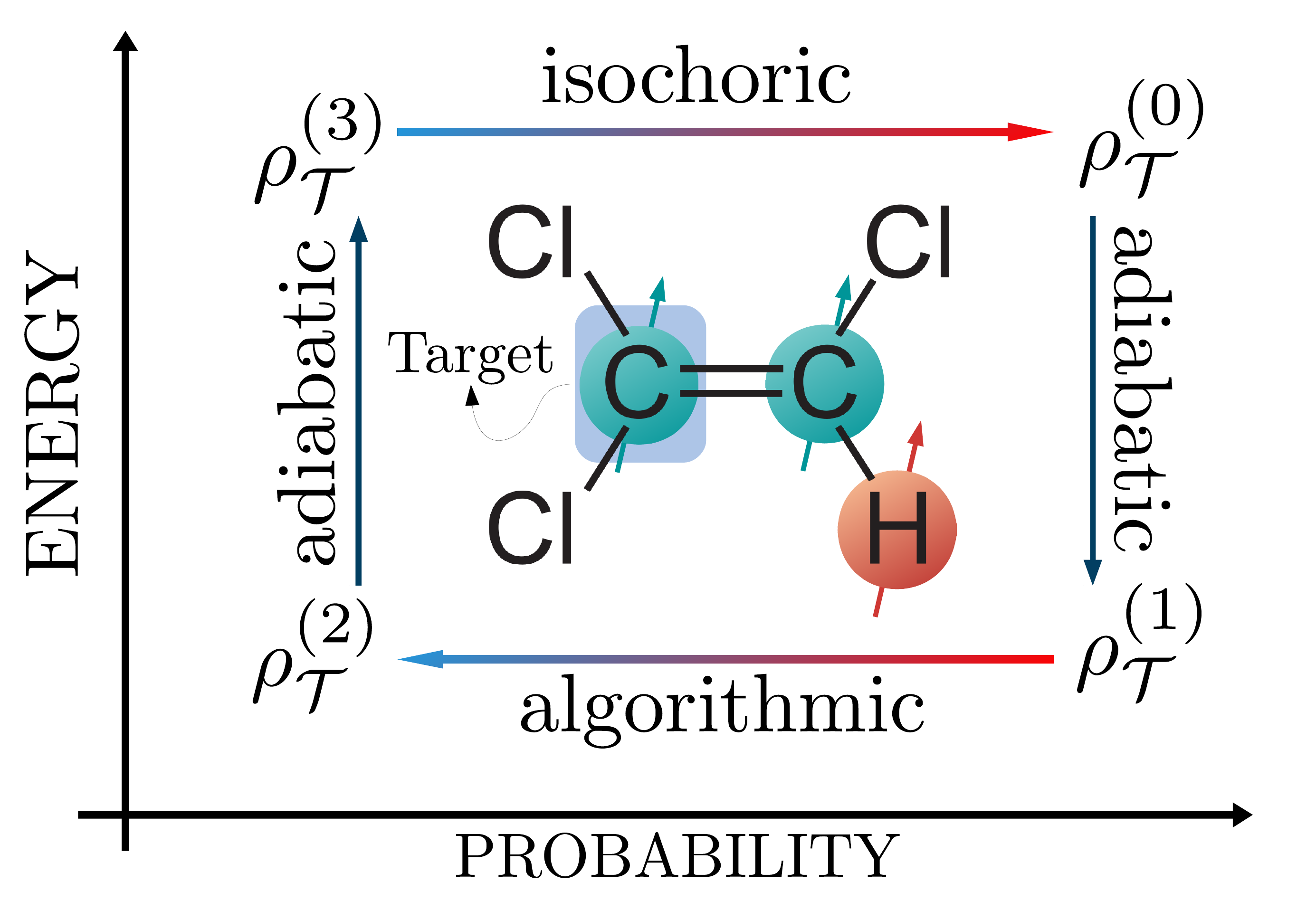}
  \caption{(Color online) Four-stroke quantum heat engine operating in a single heat bath at temperature $T$. It has one isochoric heating process, two adiabatic processes and one algorithmic cooling process.}
  \label{fig:QOE}
\end{figure}

\begin{figure}[h!]
  \includegraphics[width=0.85\linewidth]{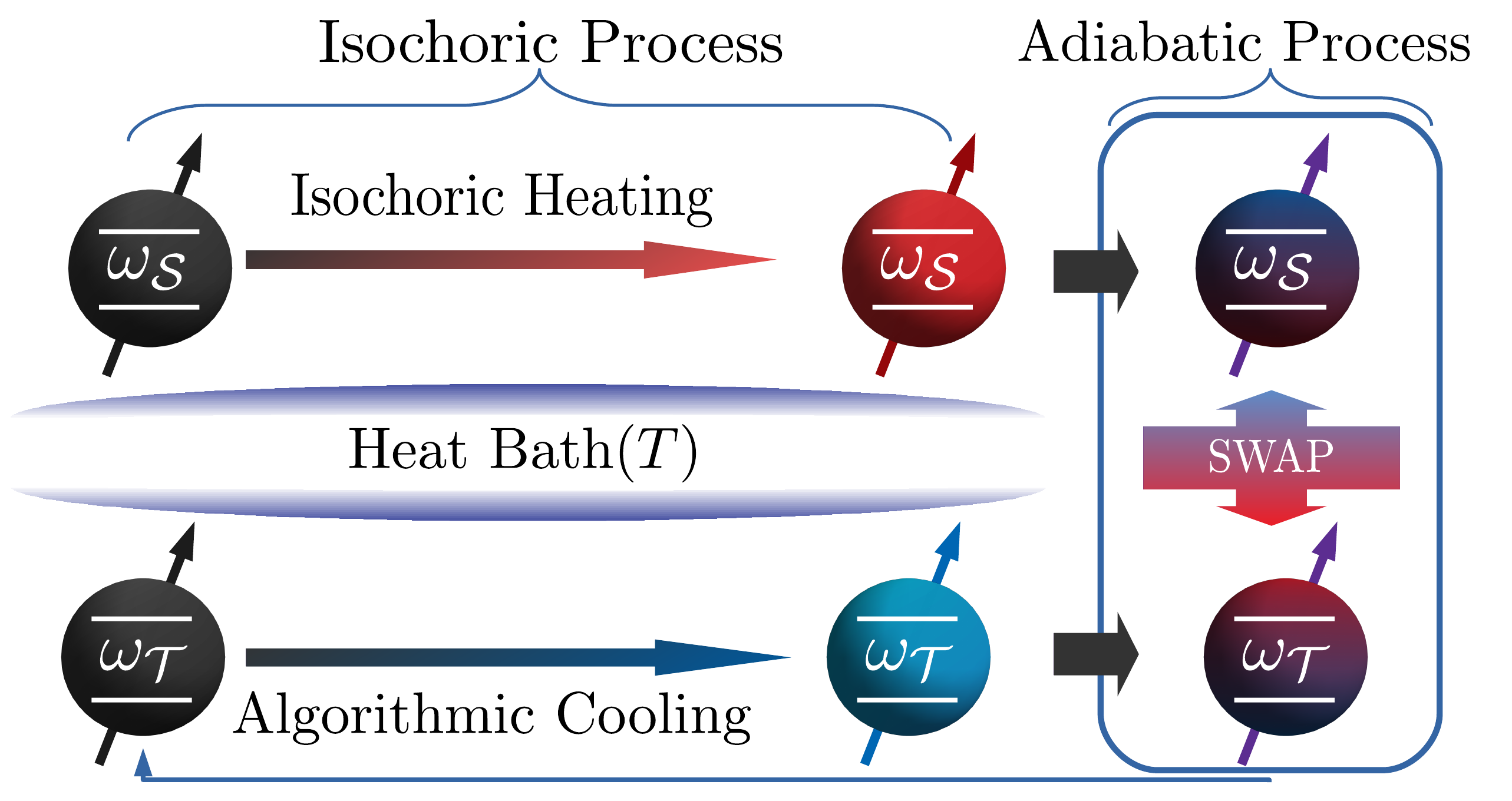}
  \caption{(Color online) Two-stroke quantum heat engine operating in a single heat bath at temperature $T$. It has  isochoric heating process and algorithmic cooling process happen at the same time. Then, one adiabatic process using SWAP operation.}
  \label{fig:QOE2}
\end{figure}

Quantum Otto engines (QOEs) consist of quantum adiabatic and isochoric processes~\cite{Cakmak2017}. Three types of quantum Otto cycle is given in the literature as four-stroke, two-stroke, and continuous~\cite{PhysRevE.76.031105,PhysRevX.5.031044}. In this article, four-stroke, and two-stroke type of QOEs are examined. The basic model of HBAC with PPA is considered instead of isochoric cooling. Implementation of HBAC requires two sets of qubits; reset qubits and computational qubits. One of the computational qubits operated as target qubit which is going to be cooled by applying PPA while other computational qubits play a role in entropy compression~\cite{PhysRevLett.116.170501}. This cooling process can be implemented with a minimal system composed of just 3-qubit~\cite{PhysRevA.93.012325, Brassard20144,Park2016,PhysRevLett.116.170501}. For the four-stroke engine, we consider the target qubit as the working fluid (see Fig.~\ref{fig:QOE}). 
As the working substance for two-stroke QOEs, we again chose the target qubit of the 3-qubit system, but with an extra qubit coupled to the target qubit (see Fig.~\ref{fig:QOE2}). 

In NMR systems, various 3-qubit models have been used for quantum information processing~\cite{Hou2014, Li2011, PhysRevA.74.062317, PhysRevLett.88.187901}. To make an applicable and realistic model for QOEs using these systems, a suitable sample and a well-designed procedure must be considered. We used parameters of $^{13} \mathrm{C_2}$-trichloroethylene (TCE) (see Table~\ref{tab:table1}) with paramagnetic reagent $\mathrm{Cr(acac)_3}$ in cloroform-d solution ($\mathrm{CDCl_3}$) in our numerical calculations, which are also experimentally used for HBAC in Refs.~\cite{PhysRevA.93.012325, Brassard20144}. Carbon-1 and Carbon-2 qubits are classified as the target and the compression qubits. And, Hydrogen qubit is selected as the reset qubit because of the relaxation time, which is small compared to the other two qubits. To implement HBAC, the Hamiltonian of 3-qubit in the lab frame can be written as~\cite{OLIVEIRA2007137}
\begin{equation}
H^{(0)} = -\hbar \sum_{i} \omega_i I_{iz} + \hbar \sum_{i \neq j}  J_{ij}I_{iz}I_{jz} ,\quad i,j=\{ \mathcal{T},\mathcal{C},\mathcal{R}\},
\label{eq1}
\end{equation}
where, $ \omega_i = \gamma_i B_z $ is the characteristic frequency of the $ i^{th} $ qubit, $\gamma _i $ is the nuclear gyromagnetic ratio and $B_z$ is the magnetic field. $\mathcal{T}$, $\mathcal{C}$ and $\mathcal{R}$ stand for target, compression and reset qubits, respectively. $I_z$ is the component of the spin angular momentum of the $i^{th}$ qubit and $J_{ij}$ is the scalar isotropic coupling strength between $i^{th}$ and $j^{th}$ qubits.
\begin{table}[h!]
  \begin{flushleft}
    \begin{tabular}{|l|c|c|c|} 
      \hline
      \textbf{} & Target-\textbf{C1} & Compression-\textbf{C2} & Reset-\textbf{H} \\
      \hline
      $\gamma/2\pi $ & 10.7084 [MHz/T] & 10.7084 [MHz/T] & 42.477 [MHz/T]\\
      \hline
      $\omega/2\pi $ & 125.77 [MHz] & 125.77 [MHz] & 500.13 [MHz]\\
      \hline
      $\tau ^1 $ & 43 [s]  & 20 [s] & 3.5 [s]\\
      \hline
      \hline
      \textbf{} & \textbf{C1-C2} & \textbf{C1-H} & \textbf{C2-H} \\
      \hline
      $J/2\pi$ & 103 [Hz] & 9 [Hz] & 200.8 [Hz]\\
      \hline
    \end{tabular}
  \end{flushleft}
  \caption{The first row in the table shows gyromagnetic ratio values of Carbon1, Carbon2 and Hydrogen qubits. Considering the 500Mhz NMR device, the corresponding characteristic frequencies are given in the second row. Also, their experimental $\tau ^1$ relaxiation times are given in the third row.  Last row shows J-coupling strenght between these qubits \cite{PhysRevA.93.012325, Brassard20144}. }
  \label{tab:table1}
\end{table}
\section{RESULTS AND DISCUSSION}
\label{sec3}
\subsection{Four-stroke Heat Bath Algoritmic Cooled Quantum Otto Engine}
\label{sec3a}
Normally, four-stroke QOEs consist of two isochoric and two adiabatic stages. However, we consider one isochoric stage, one algorithmic cooling stage, and two adiabatic stages (see Fig.~\ref{fig:QOE}). The details of the four stroke cycle is described as follows.


\textit{Isochoric Heating:} 3-qubit of TCE molecule with Hamiltonian in Eq.~(\ref{eq1}) is in contact with a heat bath at temperature T=300 K. The density matrix of the 3-qubit system at the end of this stage is given by 
\begin{equation}
\rho_{\mathrm{th}}^{(0)} = \frac{e^{-\beta H^{(0)}}}{Z}.
\label{eq11}
\end{equation}
Here $Z=\mathrm{Tr}\left[ e^{-\beta H^{(0)}}\right] $ is the partition function and $\beta = 1/k_BT $. The initial density matrix of our working fluid can be expressed by taking a partial trace of 3-qubit system
\begin{equation}
  \rho_\mathcal{T} ^{(0)}=\mathrm{Tr}_{{\mathcal{C,R}}}\left[ \rho_{\mathrm{th}} ^{(0)}\right].
\label{eq2}
\end{equation}

\textit{Adiabatic Compression:} Qubits are isolated from the heat bath and undergo finite-time adiabatic expansion. The adiabatic processes of the cycle are assumed to be generated by a time-dependent magnetic field~\cite{e15062100,doi:10.1080/00107514.2016.1201896}. Here, $H^{(0)}$ at $t=0 $ is changed to $H^{(1)}$ at $t=\tau / 2 $ by driving the initial magnetic field as $B_z\rightarrow B_z/2$. The time evolution of the density matrix is governed by the Liouville-von Neumann equation $\dot{\rho}(t) = -[H(t),\rho(t)]  $, and $H(t)$ can be expressed as $H(t) = H^{(0)} + H_{\mathrm{drive}} $, where $H_{\mathrm{drive}} $ is given by
\begin{equation}
H_\mathrm{{drive}} = \hbar \sum_{i} \left( \omega_i - \omega_i'\right) I_{iz}\sin \left(\frac{\pi t}{\tau}\right).
\end{equation}
Here, $\omega_i' = \gamma _i B_z/2$ is the characteristic frequency at the end of the adiabatic stage. Up to this point, we used Hamiltonian in Eq~(\ref{eq1}) which is written for 3-qubit. However, to find the work done in this stage, we need to consider only the target qubit. The local Hamiltonian for target qubit before adiabatic compression can be written as $ H^{(0)}_{\mathcal{T}}=-\hbar \omega_\mathcal{T} I_z $. After the adiabatic compression, it will be $ H^{(1)}_{\mathcal{T}}=-\hbar \omega_\mathcal{T}' I_z $. The initial density matrix of the target qubit $(t= 0)$ is given in Eq~(\ref{eq2}). The final density matrix of 3-qubit system at the end of adiabatic compression is $\rho^{(1)} = \rho(\tau /2)$. The density matrix of the target qubit at the end of this process is $\rho_\mathcal{T} ^{(1)}=\mathrm{Tr}_{{ \mathcal{C,R}}}\left[ \rho ^{(1)}\right]$. Then, the work performed by the working fluid is
\begin{equation}
W_1 = \mathrm{Tr} \left[ H^{(0)}_\mathcal{T} \rho_\mathcal{T} ^{(0)} \right]  - \mathrm{Tr} \left[H^{(1)}_\mathcal{T} \rho_\mathcal{T} ^{(1)}\right].
\end{equation}

\begin{figure}[h!]
  \includegraphics[width=0.9\linewidth]{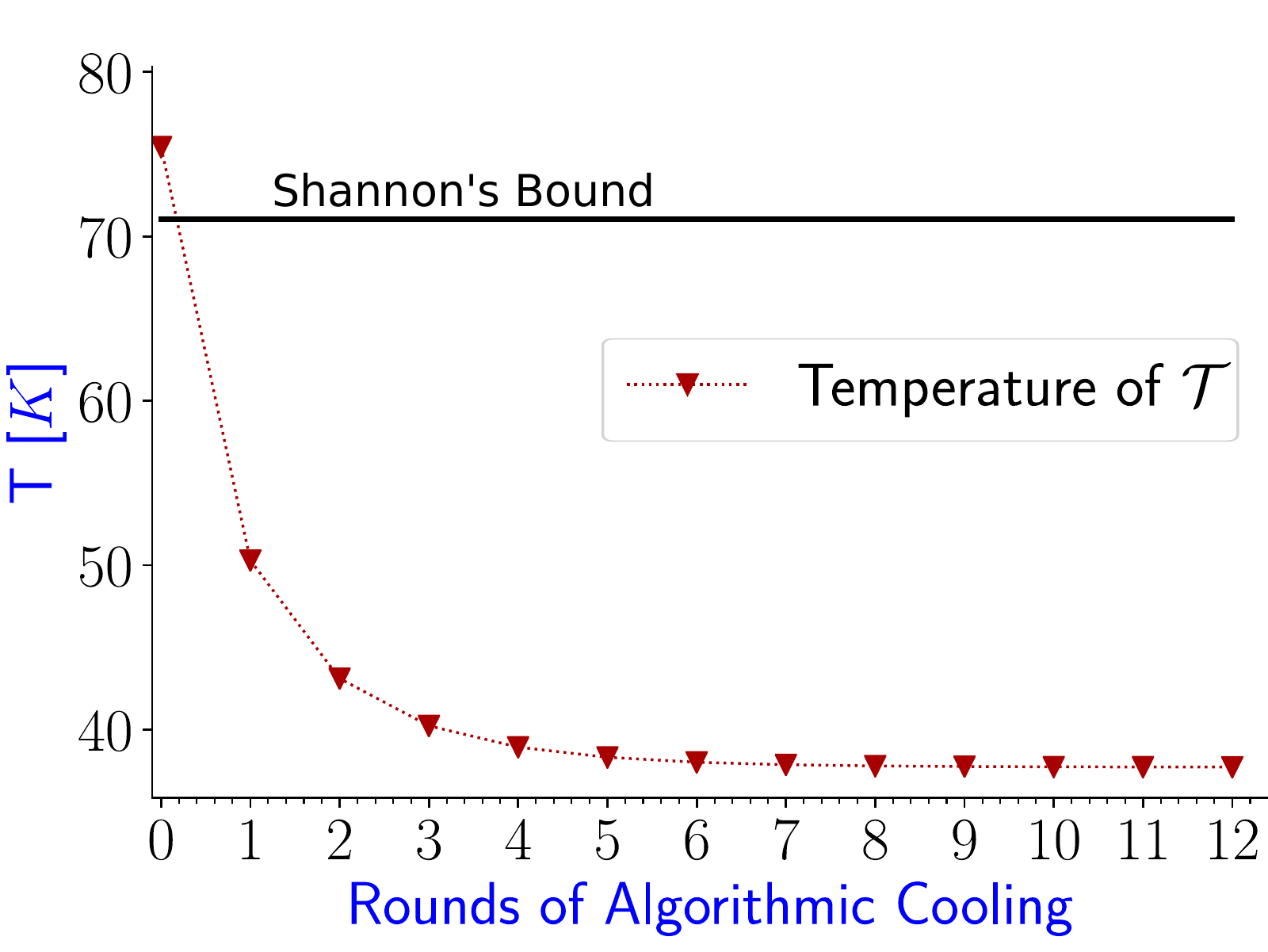} 
  \caption{(Color online) The effective temperature of target qubit calculated by using the relation between temperature and polarization given in Eq.~\ref{eq4}. The black line shows the corresponding Shannon's limit of temperature. The effective temperature of the target qubit has exceeded this limit after the first iteration.}
  \label{fig:Tem}
\end{figure}

\textit{Heat Bath Algorithmic Cooling:}
In this part of the engine, we normally need a cold heat bath to cool down our target qubit. Instead of using a cold heat bath, the working fluid treated in the same heat bath as isochoric heating. Thus, to cool down the target qubit HBAC is used. Details of the cooling mechanism given in the Appendix~\ref{apx}. For each qubit, polarization is defined as 
\begin{equation}
  \epsilon_i = P^\uparrow _i - P^\downarrow _i = \tanh \left(\frac{\hbar \gamma _i B_z}{2k_B T} \right),
\end{equation}
where, $ P^\uparrow _i$ and $ P^\downarrow _i$ denote the probability of up and down states. In a closed quantum system, Shannon's bound limits the polarization of single spin in a collection of equilibrium spin system. Using HBAC take advantage of the heat bath to cool the target qubit beyond Shannon's bound~\cite{PhysRevA.93.012325}. As a result, the polarization of the target qubit using HBAC becomes higher than the polarization of the heat bath. 
\begin{figure}[h!]
  \includegraphics[width=0.9\linewidth]{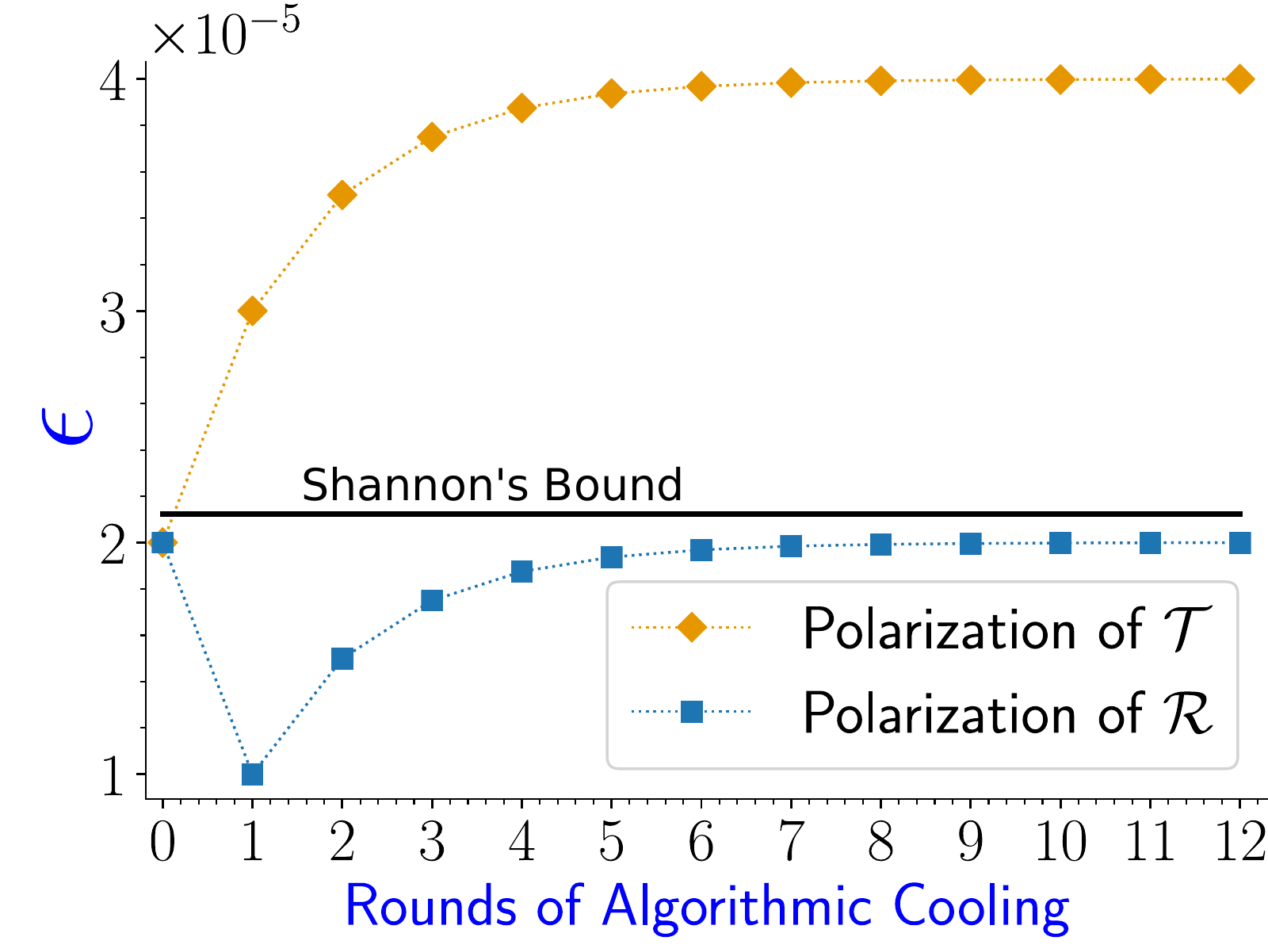} 
  \caption{(Color online) The polarization $\epsilon $ (dimensionless) calculated by using the Eq.~(\ref{eq4}) for each rounds of algorithmic cooling. The target and reset qubits polarizations calculated by taking into account the perfectly applied quantum logic gates in terms of several rounds of the PPA. The black line shows Shannon's limit of polarization. The polarization of the target qubit has exceeded this limit after the first iteration and the reset qubit stays under this limit.}
  \label{fig:POL}
\end{figure}
After the first SWAP operation before PPA, polarizations of the qubits are equal to each other. The value of their polarization is given at zeroth iteration in Fig.~\ref{fig:POL} as $ \sim 2.0 \times 10^{-5} $. Then, several rounds of PPA are applied. The effective temperature of the target qubit is determined by Eq.~(\ref{eq4}) and plotted in Fig.~\ref{fig:Tem}. The target qubit reached a polarization above the Shannon limit as a result of one round of PPA and target qubit cooled down to $\sim 50 \mathrm{K} $. After seven rounds of PPA, it almost reached to its maximum value as $ \sim 4.0 \times 10^{-5}$ and cool down to $\sim 37 \mathrm{K}$ temperature. At the end of PPA, density matrix of target qubit is given by  Eq.~(\ref{apxeq}) as
\begin{equation}
\rho^{(2)}_{\mathcal{T}}= \mathrm{Tr}_{\mathcal{C,R}} \left[ \rho^{(1,5)}_{AC} \right].
\label{eq4}
\end{equation}

\textit{Adiabatic Expansion:} In this process $H^{(1)} $ at $t=0$ is changed to $H^{(0)}$ at $t= \tau/2 $ by driving back the magnetic field as $ B_z/2 \rightarrow B_z $. The work performed by the target qubit in this process can be written as
\begin{equation}
  W_2 = \mathrm{Tr} \left[H^{(1)}_{\mathcal{T}}\rho^{(2)}_{\mathcal{T}} \right] - \mathrm{Tr} \left[H^{(0)}_{\mathcal{T}}\rho^{(3)}_{\mathcal{T}} \right].
\end{equation}

\begin{figure}[h!]
  \includegraphics[width=0.9\linewidth]{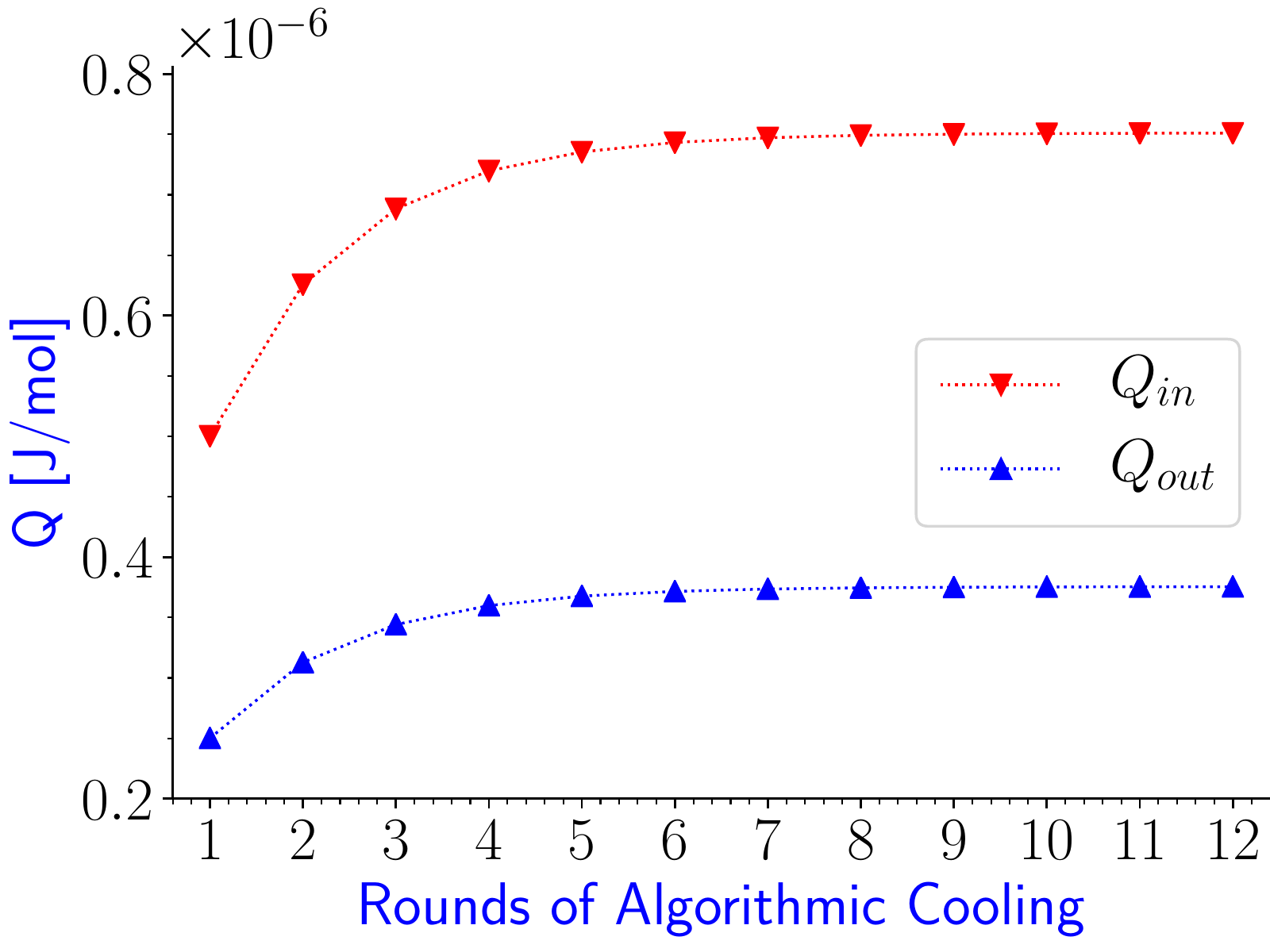}
  \caption{(Color online) Heat absorbed and released in a four-stroke cycle by the target qubit in the isochoric process ($Q_{in}$) and algorithmic cooling process ($Q_{out}$) for per rounds of PPA.}
  \label{fig:Qz}
\end{figure}

\begin{figure}[h!]
  \includegraphics[width=0.9\linewidth]{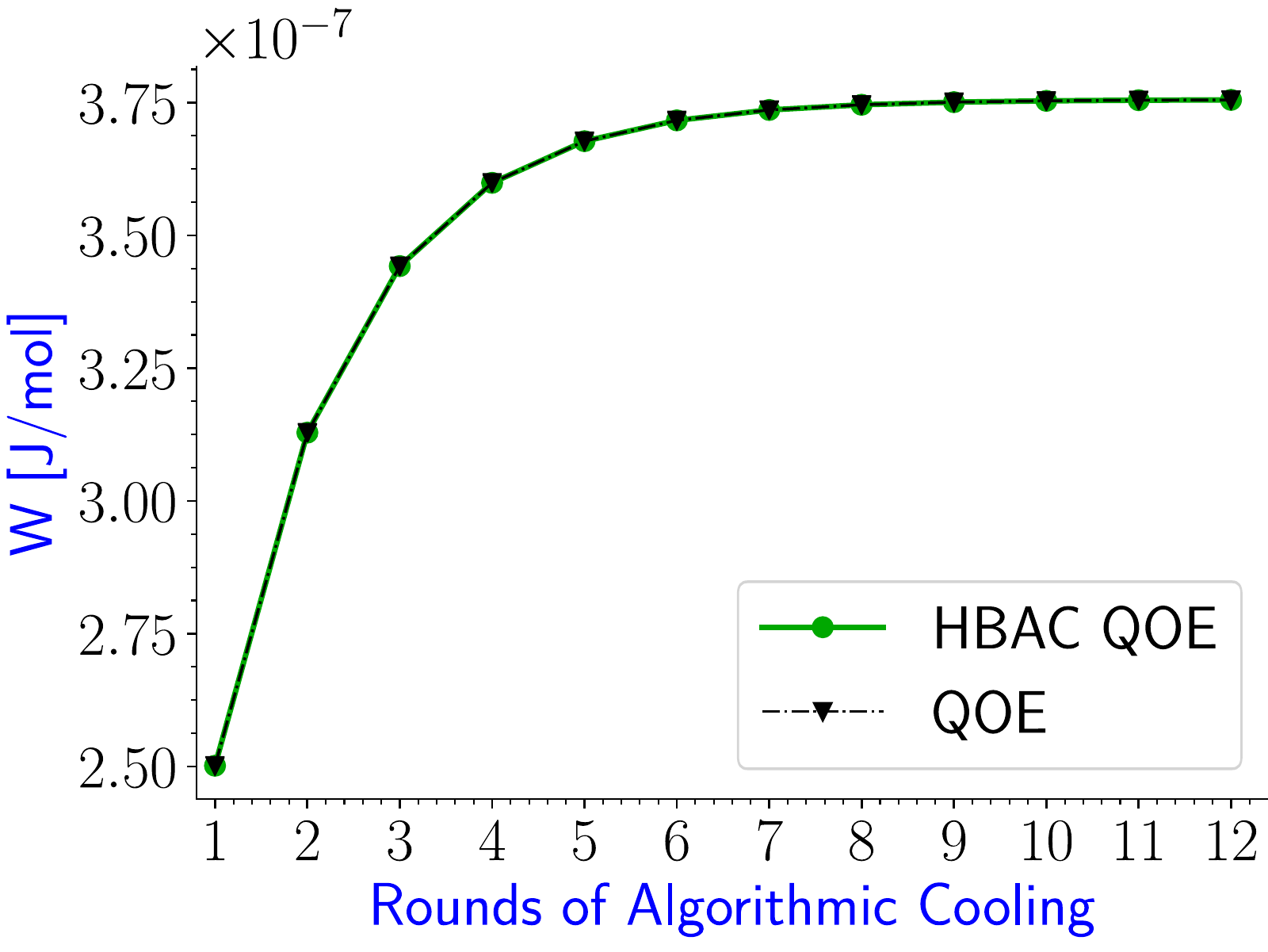}  
  \caption{(Color online) Work obtained, in a four-stroke cycle per number of iteration of PPA, from target qubits of one mol of TCE cooled by HBAC (green line). And work obtained from a mol of qubits, which are cooled by isochoric stage to temperatures corresponding in Fig~\ref{fig:Tem} (black line).}
  \label{fig:W}
\end{figure}

\begin{figure}[h!]
  \includegraphics[width=0.9\linewidth]{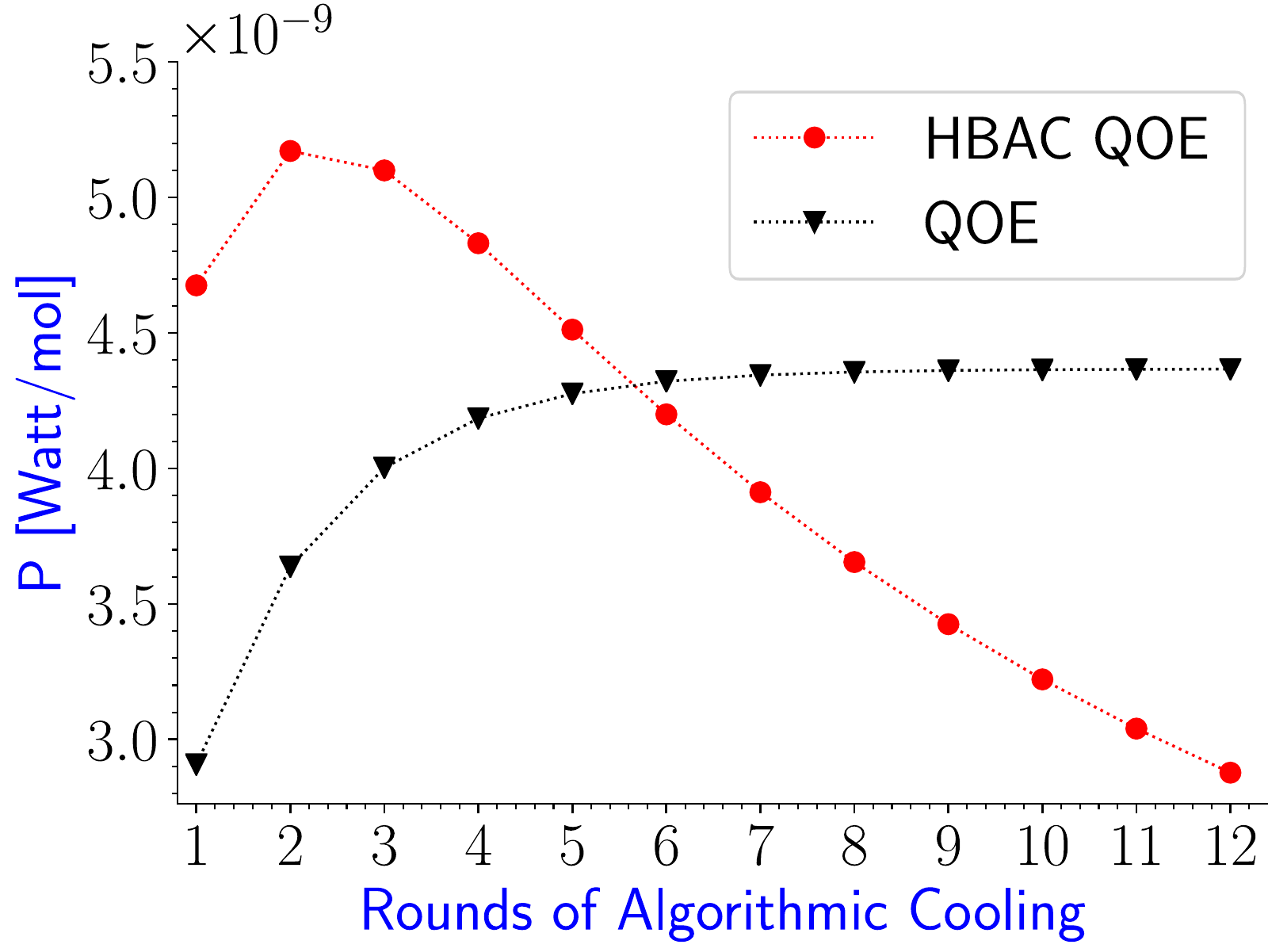}  
  \caption{(Color online) Power output for a four-stroke cycle per number of iteration of PPA, from target qubit of one mol TCE, cooled by HBAC (red) and from a mol of qubits cooled by isochoric stage to temperatures corresponding in Fig~\ref{fig:Tem} (black).}
  \label{fig:P}
\end{figure}

\textit{Work and Power Output of Four-Stroke QOE:} The total work done by the working fluid at the end of adiabatic stages can be found as $ W=W_1+W_2 $. Alternatively, total work can also be calculated from the isochoric stage and algorithmic cooling stage using $W=Q_{\text{in}}-Q_{\text{out}} $ where, $
   Q_{\text{in}} =\mathrm{Tr} \left[H^{(0)}_{\mathcal{T}} (\rho^{(3)}_\mathcal{T} -\rho^{(0)}_\mathcal{T}  )\right]$  and $
  Q_{\text{out}} =\mathrm{Tr} \left[H^{(1)}_{\mathcal{T}} (\rho^{(2)}_\mathcal{T} - \rho^{(1)}_\mathcal{T}) \right] $ are the heat released and absorbed in these stages, respectively. The efficiency of the cycle is determined by $\eta = 1-\omega_\mathcal{T}'/\omega_\mathcal{T} $ and for this engine $\eta = 0.5 $. In Fig.~\ref{fig:Qz} , we plot the heat released and absorbed by the target qubit as per rounds of PPA. As the number of iterations increases, it can be observed that the absorbed heat increases more than the heat released. After the first iteration, the target qubit absorbed $\sim 5\times 10^{-7}\mathrm{J/mol} $  of heat and released $\sim 2.5\times 10^{-7}\mathrm{J/mol} $ of heat. As a result $\sim 2.5\times 10^{-7}\mathrm{J/mol} $ work was performed. 
In order to see the difference in power caused by the algorithmic cooling and isochoric processes, we consider cold baths, corresponding to the temperatures in Fig.~\ref{fig:Tem}, to simulate a quantum Otto cycle cooled by isochoric stage with same parameters. By this way, the work output of the quantum Otto cycle cooled by isochoric stage will be the same as the cycle cooled by HBAC (see Fig~\ref{fig:W}). As we can see from Fig.~\ref{fig:W}; while the work produced by the target qubit rapidly increases with the number of iterations at the beginning, it remains constant after a certain iteration of the PPA. The reason for this behavior, the HBAC is able to cool the target qubit up to a certain limit. In Fig.~\ref{fig:Qz} , after the fourth iteration, the target qubit almost reaches the maximum value it can absorb and release heat. Absorbed and released heat from qubit in the cycle at this iteration is $\sim 3.5\times 10^{-7}\mathrm{J/mol} $ and $\sim 7.2\times 10^{-7}\mathrm{J/mol} $. Then, maximum work output of the cycle is $\sim 3.7\times 10^{-7}\mathrm{J/mol} $ (see Fig.~\ref{fig:W}). 
The number of iteration of PPA is important for quantum heat engines. Because more iteration means that more relaxation of reset qubit and it increases the time required to complete engine cycle. Even if we increase the work output iterating more PPA, we may lose power output. For a quantum Otto cycle using NMR system as working fluid, the adiabatic stages of the cycle are considered as fast compared to the isochoric stages~\cite{Cakmak2017}. We estimate the power output of cycles considering isochoric stages and HBAC. A single number of iteration of PPA requires two reset process. Taking the number of iteration 'n', we can write power output for quantum Otto cycle using HBAC as $
  P = {W}/{(\tau_\mathcal{T}+\tau_\mathcal{R}(2n+1))},
  \label{pow}
$
where, $\tau_\mathcal{T}$ and $\tau_\mathcal{R} $ are relaxation times of the Carbon1 and the Hydrogen qubits respectively from the Table ~\ref{tab:table1}. For the cycle using isochoric process in the cooling stage, power output is $P = W/2\tau_\mathcal{T} $. In Fig.~\ref{fig:P}, we plot power output for these engines. We see that the second iteration of PPA gives maximum power output as $\sim 5.2\times10^{-9} $ Watt/mol. For more number of iteration, this power output is getting decrease. 
After the fifth iteration of PPA, the cycle using the isochoric cooling stage can dominate the engine using the HBAC stage. Accordingly, the optimum choice of the number of rounds in HBAC is two for our four-stroke model system. Such a choice optimizes the power output of the cycle yielding high power performance. 
\subsection{Two-Stroke Heat-Bath Algorithmic Cooled Quantum Otto Engine}
\label{sec3b}
\begin{figure}[h!]
  \includegraphics[width=0.99\linewidth]{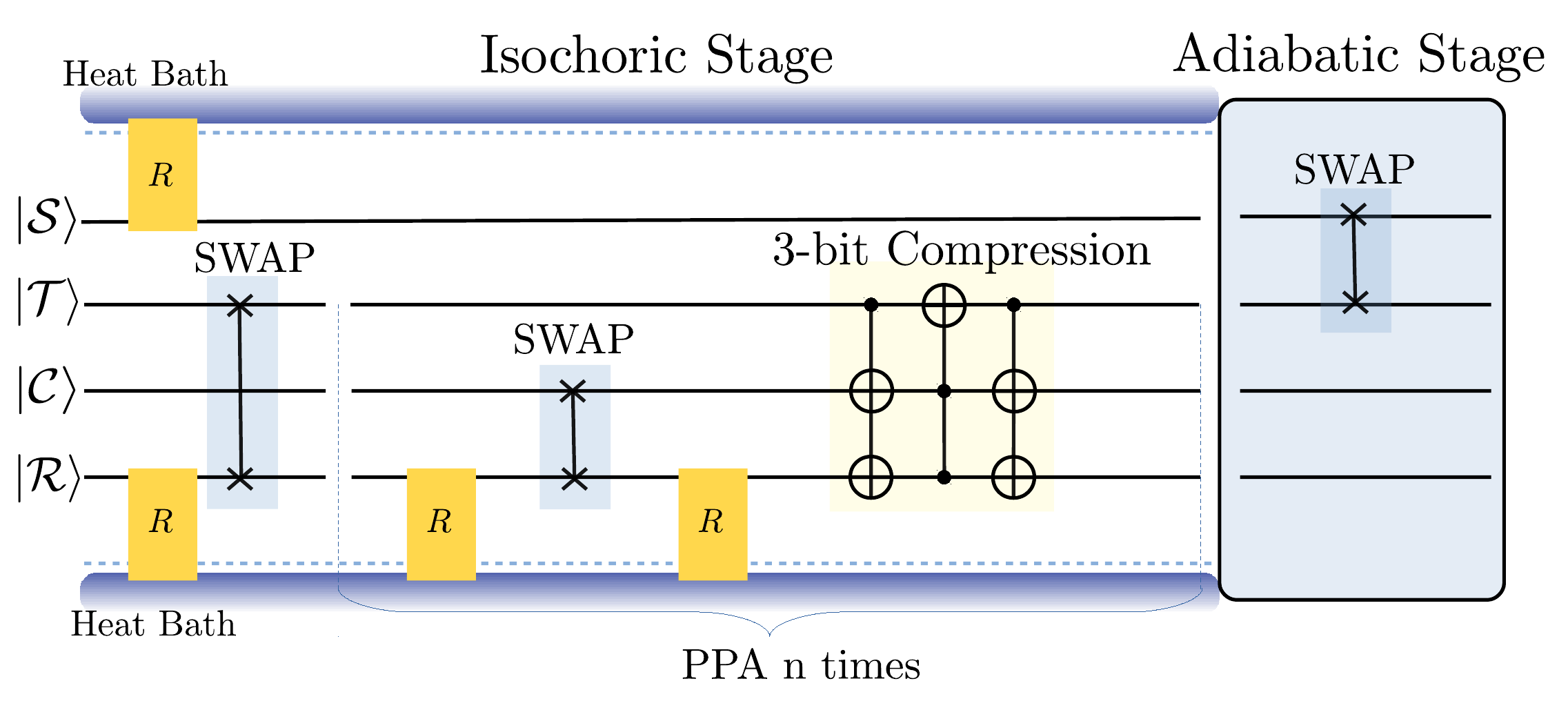}
  \caption{Quantum circuit demonstrating the two stroke heat bath algorithmic cooled quantum Otto engine. $|\mathcal{S}\rangle$ stands for qubit $\mathcal{S} $ and $|\mathcal{T}\rangle$ for qubit $\mathcal{T}$. Here, isochoric heating demonstrated as R operation and applied only one time on qubit $ \mathcal{S} $. For qubit $ \mathcal{T}$, PPA applied and it is given in Appendix~\ref{apx}. After these two process finished SWAP operation applied between two qubits and cycle is completed with adiabatic process.}
  \label{fig:2strokesalg}
\end{figure}

\begin{figure}[h!]
  \includegraphics[width=0.9\linewidth]{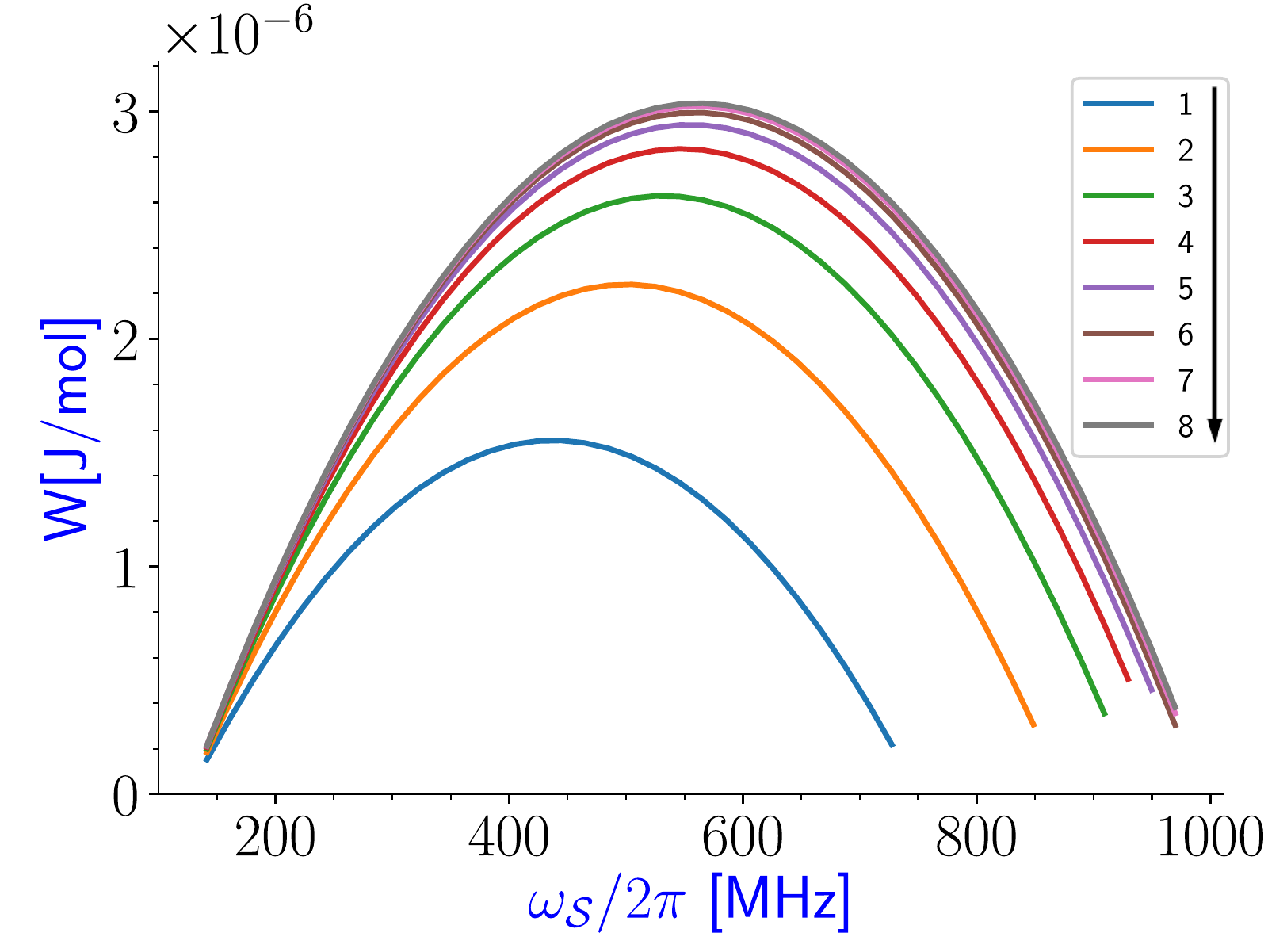}
  \caption{(Color online) Positive work obtained in 2 strokes HBAC-QOE as a function of $\omega_\mathcal{S}$. The legend of the plot shows the number of rounds of PPA applied to qubit $\mathcal{T}$ and direction of the arrow shows an increase in work output from 1st iteration to 8th iteration.}
  \label{work2}
\end{figure}

\begin{figure}[h!]
  \includegraphics[width=0.9\linewidth]{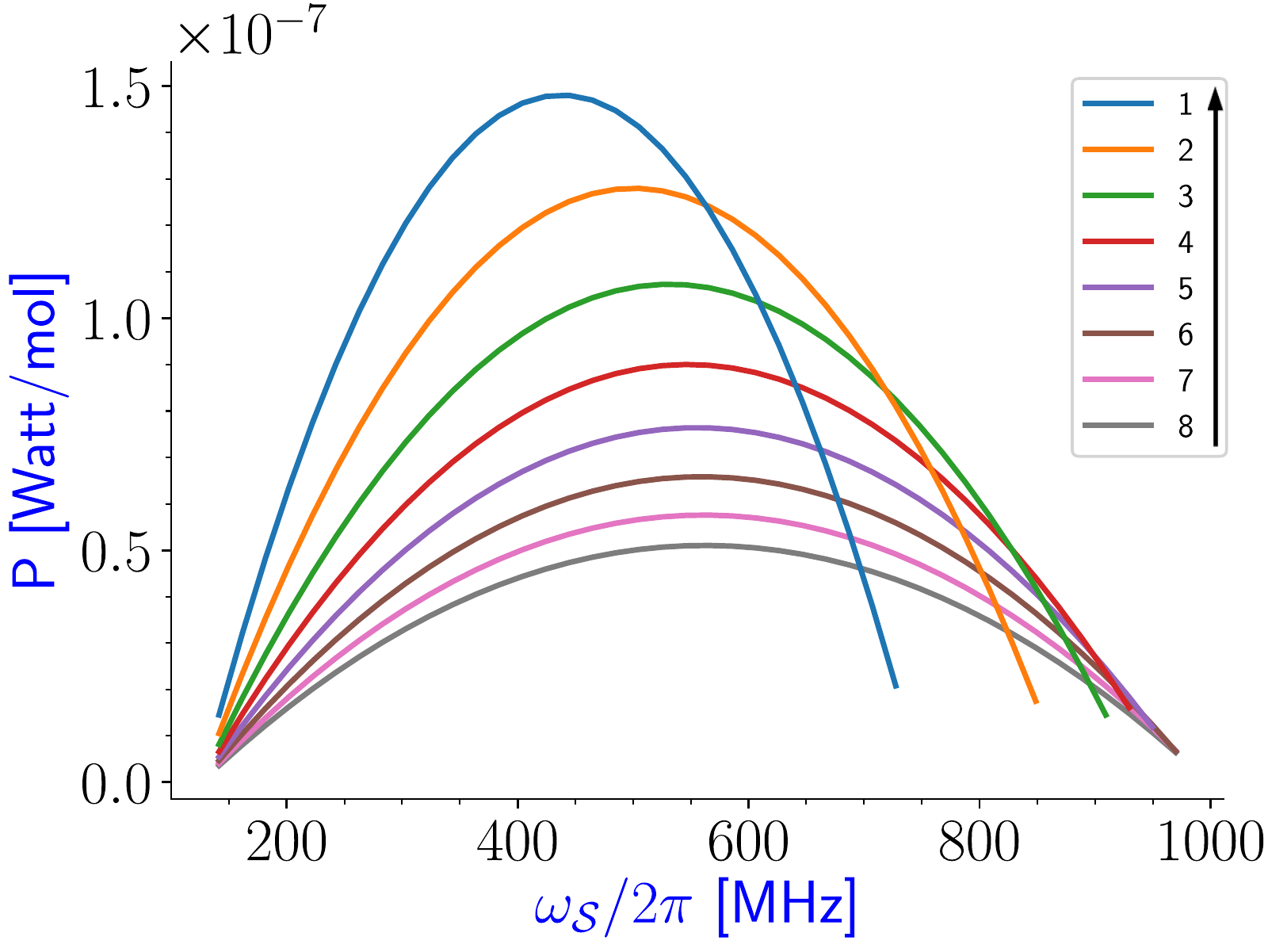}
  \caption{(Color online) Power output obtained in a two-stroke quantum Otto cycle for the different number of iterations of PPA as a function of  $\omega_\mathcal{S} $. The legend of the plot shows the number of rounds of PPA applied to qubit $\mathcal{T}$ and the direction of the arrow shows an increase in the power output from 8th iteration to 1st iteration.} 
  \label{power}
\end{figure}

We also investigate two-stroke QOE that is proposed in Refs. \cite{PhysRevE.76.031105, PhysRevX.5.031044}. To construct it, we need to consider two qubits donated as $\mathcal{S}$ and $\mathcal{T}$ as the working fluid. First, two qubits are isolated from each other. One of the qubits contacts with a heat bath at temperature T until it reaches to equilibrium. Our purpose is to cool the other qubit within the same heat bath utilizing algorithmic cooling. Hence we need another two qubits for this process as compression qubit and reset qubit. This part is considered as two isochoric processes for the heat engine. Second, two qubits decouple from the heat bath. Then, the SWAP operation is performed between these two qubits (see Fig.~\ref{fig:2strokesalg}).
Local Hamiltonian of the qubit $ \mathcal{S} $ and qubit $\mathcal{T} $ can be written as $H_\mathcal{S} = -\hbar \omega_\mathcal{S} I_z$ and $H_\mathcal{T} = -\hbar \omega_\mathcal{T} I_z$, respectively. And density matrix of qubit $\mathcal{S}$, at the end of the isochoric stage is given by $\rho^{(0)}_\mathcal{S} =e^{-\beta H_{\mathcal{S}}}/{Z}$. Using the PPA given in the Appendix~\ref{apx}, we can write the density matrix of qubit $\mathcal{T}$ at the end of HBAC as $\rho^{(0)}_\mathcal{T}=\mathrm{Tr}_{\mathcal{C,R}} \left[ \rho^{(1,5)}_{AC} \right]$. It is assumed that the coupling between $\mathcal{S} $ and $\mathcal{T} $ qubits are small compered to $\omega_\mathcal{S}$ and $\omega_\mathcal{T}$. The density matrix of the total working fluid can be expressed as
\begin{equation}
  \rho_{\mathcal{S,T}}^{(0)} \approx  \rho^{(0)}_\mathcal{S} \otimes \rho^{(0)}_\mathcal{T}.
\end{equation}
In the adiabatic process, a SWAP gate is applied to exchange the states of $\mathcal{S}$ and $\mathcal{T} $ qubits
\begin{equation}
  \rho_{\mathcal{S,T}}^{(1)} = \mathrm{SWAP}\left( \rho_{\mathcal{S,T}}^{(0)} \right) \mathrm{SWAP} ^\dagger.
\end{equation}
At the end of the cycle, density matrices of individual qubits become
\begin{equation}
  \rho^{(1)}_\mathcal{S} =\mathrm{Tr}_\mathcal{T} \left[\rho^{(1)}_\mathcal{S,T} \right],\quad  \rho^{(1)}_\mathcal{T} =\mathrm{Tr}_\mathcal{S} \left[\rho^{(1)}_\mathcal{S,T} \right].
\end{equation}
Heat absorbed by the qubit $\mathcal{S}$ calculated as follows
\begin{equation}
  Q_{in} =\mathrm{Tr} \left[H_\mathcal{S} \rho^{(0)}_\mathcal{S} \right] - \mathrm{Tr} \left[H_\mathcal{S} \rho^{(1)}_\mathcal{S} \right],
\end{equation}
and the heat released by the qubit $\mathcal{T}$ is
\begin{equation}
  Q_{out} =\mathrm{Tr} \left[H_\mathcal{T} \rho^{(1)}_\mathcal{T} \right] - \mathrm{Tr} \left[H_\mathcal{T} \rho^{(0)}_\mathcal{T} \right].
\end{equation}
The net work is evaluated by  $W=Q_{\mathrm{in}}-Q_{\mathrm{out}}$, with efficiency $\eta= 1-\omega_\mathcal{T}/\omega_\mathcal{S}$. To get a positive work, the frequency of qubit $\mathcal{S} $ needs to be greater than the frequency of the qubit $\mathcal{T}$ ($\omega_\mathcal{S}>\omega_\mathcal{T}$). In addition, $ T > T_\mathcal{T} ({\omega_\mathcal{S}}/{\omega_\mathcal{T}}) $ needs to be satisfied. Here $T_\mathcal{T} $ is the temperature of target qubit in HBAC. Positive work conditions then can be expressed as
\begin{equation}
\omega_\mathcal{T} < \omega_\mathcal{S} < \omega_\mathcal{T} \frac{T}{T_\mathcal{T}}. 
\label{pwc}
\end{equation} 
Fig.~\ref{work2} shows the relation between the work output and $\omega_\mathcal{S} $, for different number of iterations. When the number of iteration of PPA is increased, the best work output is also increased. 
However, after some number of iteration, it remains almost constant, because of the limitation of PPA to cool qubit $\mathcal{T} $. It is found that for 580 MHz, we almost get the best work output at 5th and more iterations as $\sim 3.0\times 10^{-6} $ J/mol. But this frequency does not give us the best work output at 1st iteration, 430 MHz gives. For 1st iteration with 430 MHz frequency we get $\sim 1.5\times 10^{-6} $ J/mol work output.  The adiabatic stage is evaluated by a SWAP operation, which is fast compared to the HBAC. Relaxation time of the $\mathcal{S}$ qubit may be estimated by considering spectral density functions $J(\omega, t _{c}) $, where $t _c $ is the correlation time~\cite{5392713,Abragam1963}. If we look for the optimum power output, the $\omega_\mathcal{S} $ is close to the frequency of the $\mathcal{R}$ qubit. Considering environmental effects are the same, $t_c$ may be assumed to be close for the $\mathcal{R}$ and $\mathcal{S} $ qubit. As a result, we can say that the $\mathcal{S}$ qubit is thermalized until the HBAC process is complete. In addition, using isochoric cooling instead of HBAC to cool target qubit requires more time up to the 5th iteration of PPA, as we have shown in the Sec.~\ref{sec3a}. Thus, estimation of the power output depends on the relaxation time of reset qubit and the number of rounds of PPA. Then, we can write power output as $ P = {W}/{\tau_\mathcal{R}(2n+1)}$. The relaxation time of Hydrogen ($\tau_\mathcal{R} $) is given in Table ~\ref{tab:table1}. We plot the power output of the cycle in Fig.~\ref{power}. 
When the number of iterations increased, the power output is decreased despite the increase in the work output. The optimum value given by 1st iteration with 430 MHz frequency of $\mathcal{S}$ qubit as $\sim 1.47\times 10^{-7} $ Watt/mol. If we compare these results to four-stroke QOEs both cooled by HBAC and isochoric stage, we see that two stroke cycle gives more work and power output. The efficiency of the cycle as a function of $\omega_\mathcal{S} $ is given above. Taking the optimum value for the power output at 1st iteration and 430 MHz frequency of $\mathcal{S}$ qubit, the efficiency of the cycle is 0.7, which also more efficient than four-stroke QOEs.
\section{CONCLUSIONS}
\label{con}
We have investigated the possible quantum Otto engines considering HBAC instead of isochoric cooling. In conventional NMR setups do not let to change strength (huge) magnetic field along the z direction. In order to solve this restriction, NMR setup can be modified to change strong magnetic field via gradient coils such as Magnetic Resonance Imaging (MRI) system. In addition, the sample always in a single entropy sink in NMR systems. This is the main problem to design QOEs, which requires two heat bath to extract work from NMR qubits. Using HBAC allowed us to cool the working fluid in a single heat bath. Here we specifically showed this cooling process can be implemented to four-stroke and two-stroke QOEs. The isochoric cooling process of the cycle takes too much time compared to the HBAC. Comparing our results with a single spin NMR heat engine, utilizing the HBAC to QOEs improves the power output of the cycles up to a certain iteration of PPA.  
\acknowledgements
The authors thank to M.~Paternostro for fruitful discussions.

\appendix
\section{}
\label{apx}  
\subsection{Heat Bath Algorithmic Cooling - Partner Pairing Algorithm(PPA)}

\begin{figure}[h!]
  \includegraphics[width=0.99\linewidth]{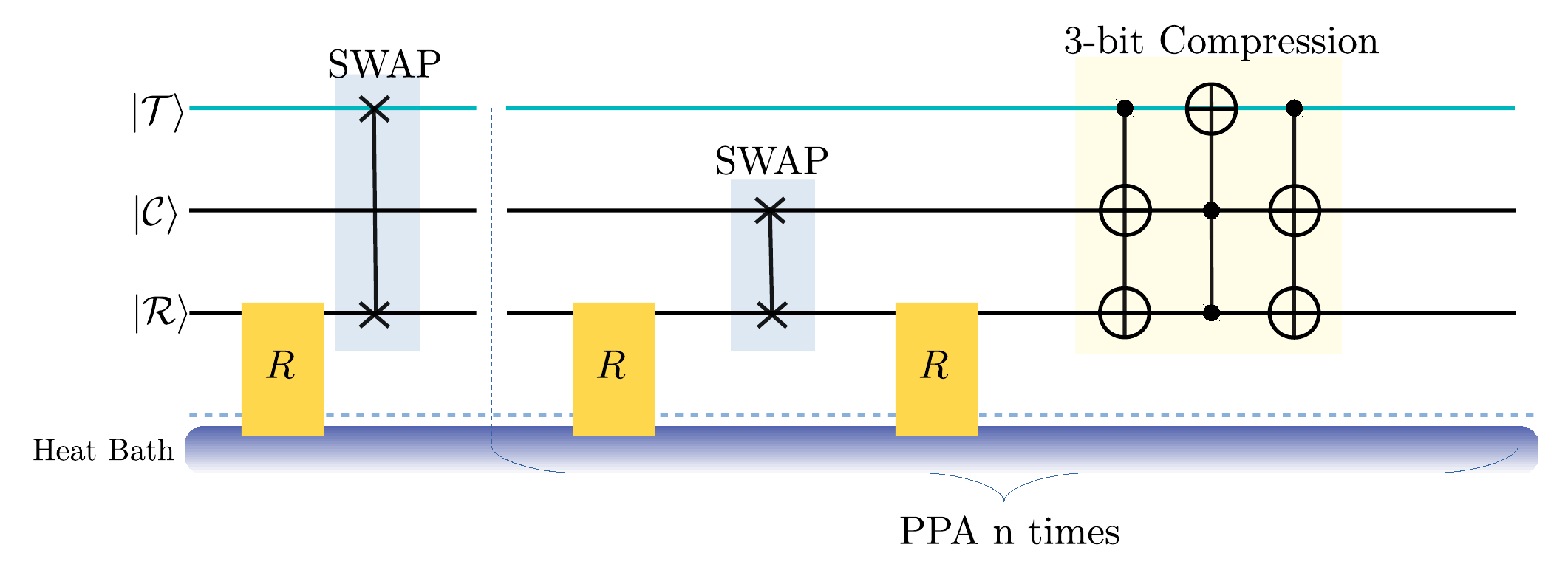}
  \caption{Quantum circuit demonstating partner pairing algorithm for 3-qubit system given in Ref.~\cite{PhysRevLett.116.170501}. $|\mathcal{T}\rangle$, $|\mathcal{C}\rangle$ and $|\mathcal{R}\rangle$ stands for target, compression  and reset qubits respectively. R process means the relaxation of reset qubit. In PPA, the first reset process applied only one time. Then, the iteration part consists of SWAP and 3-bit compression operations applied n times.}
  \label{fig:AC}
\end{figure}

We consider heat bath algorithmic cooling (HBAC) with partner pairing algorithm (PPA) (see Fig.~\ref{fig:AC}), which uses quantum information processing to increase the purification level of qubits in NMR systems~\cite{PhysRevLett.116.170501,Park2016}.  
Before starting PPA, the individual density matrices of the target, the compression and the reset qubits are $\rho_\mathcal{T} ^{(1)}=\mathrm{Tr}_{{ \mathcal{C,R}}}\left[ \rho ^{(1)}\right] $, $\rho_\mathcal{C} ^{(1)}=\mathrm{Tr}_{{ \mathcal{T,R}}}\left[ \rho ^{(1)}\right]$,  $ \rho_\mathcal{R} ^{(1)}= \mathrm{Tr}_{{\mathcal{T,C}}}\left( \rho ^{(1)}\right) $ respectively and $\rho^{(1)} $ is the initial density matrix of 3-qubit system. The reset has small relaxation time compared to the target ($\tau _R \ll \tau _T $) and the compression ($\tau _R \ll \tau _C $) qubit, where $\tau _\mathcal{T}, \tau _\mathcal{C}$ and $ \tau _\mathcal{R}$ are given in Table \ref{tab:table1} respectively. In each step of HBAC, $R$ operation is applied to the reset qubit to thermalize it with the heat bath  $(\rho^{(1)}_\mathcal{R} \rightarrow \rho^{(1)}_{\mathcal{R},{th}} )$ at temperature T. Scalar couplings between qubits are small compared to $\omega_i$ values. If we write the total density matrix of three qubits as a tensor product of the individual states, the fidelity of the density matrix in Eq.~\ref{eq11} and this product density matrix numerically is found to be close to 1. Thus, at first step, the density matrix of 3-qubit system can be written as a tensor product of these states 
\begin{equation}
\rho^{(1,0)}_{AC} = \rho_\mathcal{T} ^{(1)} \otimes \rho_\mathcal{C} ^{(1)} \otimes \rho^{(1)}_{\mathcal{R},{th}},
\end{equation}
where, 1st index of $\rho^{(1,0)}_{AC}$ stands for stage of the cycle and 2nd index from 0 to 5 indicates the state after each process of HBAC. After the reset qubit regains its polarization, a unitary SWAP operator is used to exchange polarizations of the target and the reset qubit.
\begin{equation}
\rho^{(1,1)} _{AC} = \mathrm{SWAP}_\mathcal{T,R} \left( \rho^{(1,0)}_{AC}\right)  \mathrm{SWAP}_{\mathcal{T,R}} ^\dag
\end{equation}
The states of the target and compression qubits become $  \rho_\mathcal{T} ^{(1,1)}=\mathrm{Tr}_{{\mathcal{C,R}}}\left[ \rho^{(1,1)} _{AC} \right] $ and  $ \rho_C ^{(1,1)}=\mathrm{Tr}_{{\mathcal{T,R}}}\left[\rho^{(1,1)} _{AC}\right]$ respectively. After the unitary evolution, PPA can be applied n times, which is given as follows
\begin{enumerate}
  \item The reset qubit is thermalized with the heat bath. The density matrix of the 3-qubit system becomes
  \begin{equation}
  \rho^{(1,2)}_{AC} = \rho_\mathcal{T} ^{(1,1)} \otimes \rho_\mathcal{C} ^{(1,1)} \otimes \rho^{(1)}_{\mathcal{R},{th}}
  \end{equation}
  \item SWAP is applied to change polarization between the compression and the reset qubit, such that
  \begin{equation}
  \rho^{(1,3)} _{AC} = \mathrm{SWAP}_\mathcal{{C,R}} \left( \rho^{(1,2)}_{AC}\right)  \mathrm{SWAP}_{\mathcal{C,R}} ^\dag
  \end{equation}
  Then, the states of the target and the compression qubits becomes  $  \rho_\mathcal{T} ^{(1,3)}=\mathrm{Tr}_{{ \mathcal{C,R}}}\left[ \rho^{(1,3)} _{AC} \right] $ and  $ \rho_C ^{(1,3)}=\mathrm{Tr}_{{\mathcal{T,R}}}\left[\rho^{(1,3)} _{AC}\right]$.
  \item The reset qubit is regained its polarization. The state of 3-qubit system is expressed as
  \begin{equation}
  \rho^{(1,4)}_{AC} = \rho_\mathcal{T} ^{(1,3)} \otimes \rho_\mathcal{C} ^{(1,3)} \otimes \rho^{(1)}_{\mathcal{R},{th}}.
  \end{equation}
  \item To lower the entropy of the target qubit and to increase the entropy of the reset qubit  3-bit compression gate applied to the density matrix.
  \begin{equation}
  \rho^{(1,5)}_{AC} = \mathrm{COMP} \left( \rho^{(1,4)}_{AC} \right) \mathrm{COMP}^\dag
  \end{equation}
  where COMP is the unitary operation of compression gate composed of unitary operators as two control-not-not gates and a Toffoli gate, which can be expressed as  
  \begin{equation}
  \mathrm{COMP=[CNotNot][Toffoli][CNotNot]}.
  \end{equation}
\end{enumerate}
At the end of this iterative step the target qubit is cooled and the density matrix of it becomes
\begin{equation}
\rho^{(2)}_{\mathcal{T}}= \mathrm{Tr}_{\mathcal{C,R}} \left[ \rho^{(1,5)}_{AC} \right].
\label{apxeq}
\end{equation}

\nocite{*}

\bibliography{apssamp}

\begin{thebibliography}{57}%
\makeatletter
\providecommand \@ifxundefined [1]{%
 \@ifx{#1\undefined}
}%
\providecommand \@ifnum [1]{%
 \ifnum #1\expandafter \@firstoftwo
 \else \expandafter \@secondoftwo
 \fi
}%
\providecommand \@ifx [1]{%
 \ifx #1\expandafter \@firstoftwo
 \else \expandafter \@secondoftwo
 \fi
}%
\providecommand \natexlab [1]{#1}%
\providecommand \enquote  [1]{``#1''}%
\providecommand \bibnamefont  [1]{#1}%
\providecommand \bibfnamefont [1]{#1}%
\providecommand \citenamefont [1]{#1}%
\providecommand \href@noop [0]{\@secondoftwo}%
\providecommand \href [0]{\begingroup \@sanitize@url \@href}%
\providecommand \@href[1]{\@@startlink{#1}\@@href}%
\providecommand \@@href[1]{\endgroup#1\@@endlink}%
\providecommand \@sanitize@url [0]{\catcode `\\12\catcode `\$12\catcode
  `\&12\catcode `\#12\catcode `\^12\catcode `\_12\catcode `\%12\relax}%
\providecommand \@@startlink[1]{}%
\providecommand \@@endlink[0]{}%
\providecommand \url  [0]{\begingroup\@sanitize@url \@url }%
\providecommand \@url [1]{\endgroup\@href {#1}{\urlprefix }}%
\providecommand \urlprefix  [0]{URL }%
\providecommand \Eprint [0]{\href }%
\providecommand \doibase [0]{http://dx.doi.org/}%
\providecommand \selectlanguage [0]{\@gobble}%
\providecommand \bibinfo  [0]{\@secondoftwo}%
\providecommand \bibfield  [0]{\@secondoftwo}%
\providecommand \translation [1]{[#1]}%
\providecommand \BibitemOpen [0]{}%
\providecommand \bibitemStop [0]{}%
\providecommand \bibitemNoStop [0]{.\EOS\space}%
\providecommand \EOS [0]{\spacefactor3000\relax}%
\providecommand \BibitemShut  [1]{\csname bibitem#1\endcsname}%
\let\auto@bib@innerbib\@empty
\bibitem [{\citenamefont {Quan}\ \emph {et~al.}(2007)\citenamefont {Quan},
  \citenamefont {Liu}, \citenamefont {Sun},\ and\ \citenamefont
  {Nori}}]{PhysRevE.76.031105}%
  \BibitemOpen
  \bibfield  {author} {\bibinfo {author} {\bibfnamefont {H.~T.}\ \bibnamefont
  {Quan}}, \bibinfo {author} {\bibfnamefont {Yu-xi}\ \bibnamefont {Liu}},
  \bibinfo {author} {\bibfnamefont {C.~P.}\ \bibnamefont {Sun}}, \ and\
  \bibinfo {author} {\bibfnamefont {Franco}\ \bibnamefont {Nori}},\ }\bibfield
  {title} {\enquote {\bibinfo {title} {Quantum thermodynamic cycles and quantum
  heat engines},}\ }\href {\doibase 10.1103/PhysRevE.76.031105} {\bibfield
  {journal} {\bibinfo  {journal} {Phys. Rev. E}\ }\textbf {\bibinfo {volume}
  {76}},\ \bibinfo {pages} {031105} (\bibinfo {year} {2007})}\BibitemShut
  {NoStop}%
\bibitem [{\citenamefont {Huang}\ \emph {et~al.}(2014)\citenamefont {Huang},
  \citenamefont {Niu}, \citenamefont {Xiu},\ and\ \citenamefont
  {Yi}}]{Huang2014}%
  \BibitemOpen
  \bibfield  {author} {\bibinfo {author} {\bibfnamefont {Xiao-Li}\ \bibnamefont
  {Huang}}, \bibinfo {author} {\bibfnamefont {Xin-Ya}\ \bibnamefont {Niu}},
  \bibinfo {author} {\bibfnamefont {Xiao-Ming}\ \bibnamefont {Xiu}}, \ and\
  \bibinfo {author} {\bibfnamefont {Xue-Xi}\ \bibnamefont {Yi}},\ }\bibfield
  {title} {\enquote {\bibinfo {title} {Quantum stirling heat engine and
  refrigerator with single and coupled spin systems},}\ }\href {\doibase
  10.1140/epjd/e2013-40536-0} {\bibfield  {journal} {\bibinfo  {journal} {The
  European Physical Journal D}\ }\textbf {\bibinfo {volume} {68}},\ \bibinfo
  {pages} {32} (\bibinfo {year} {2014})}\BibitemShut {NoStop}%
\bibitem [{\citenamefont {Thomas}\ and\ \citenamefont
  {Johal}(2014)}]{Thomas2014}%
  \BibitemOpen
  \bibfield  {author} {\bibinfo {author} {\bibfnamefont {George}\ \bibnamefont
  {Thomas}}\ and\ \bibinfo {author} {\bibfnamefont {Ramandeep~S.}\ \bibnamefont
  {Johal}},\ }\bibfield  {title} {\enquote {\bibinfo {title} {Friction due to
  inhomogeneous driving of coupled spins in a quantum heat engine},}\ }\href
  {\doibase 10.1140/epjb/e2014-50231-1} {\bibfield  {journal} {\bibinfo
  {journal} {The European Physical Journal B}\ }\textbf {\bibinfo {volume}
  {87}},\ \bibinfo {pages} {166} (\bibinfo {year} {2014})}\BibitemShut
  {NoStop}%
\bibitem [{\citenamefont {Zhang}\ \emph {et~al.}(2007)\citenamefont {Zhang},
  \citenamefont {Liu}, \citenamefont {Chen},\ and\ \citenamefont
  {Li}}]{PhysRevA.75.062102}%
  \BibitemOpen
  \bibfield  {author} {\bibinfo {author} {\bibfnamefont {Ting}\ \bibnamefont
  {Zhang}}, \bibinfo {author} {\bibfnamefont {Wei-Tao}\ \bibnamefont {Liu}},
  \bibinfo {author} {\bibfnamefont {Ping-Xing}\ \bibnamefont {Chen}}, \ and\
  \bibinfo {author} {\bibfnamefont {Cheng-Zu}\ \bibnamefont {Li}},\ }\bibfield
  {title} {\enquote {\bibinfo {title} {Four-level entangled quantum heat
  engines},}\ }\href {\doibase 10.1103/PhysRevA.75.062102} {\bibfield
  {journal} {\bibinfo  {journal} {Phys. Rev. A}\ }\textbf {\bibinfo {volume}
  {75}},\ \bibinfo {pages} {062102} (\bibinfo {year} {2007})}\BibitemShut
  {NoStop}%
\bibitem [{\citenamefont {Huang}\ \emph {et~al.}(2013)\citenamefont {Huang},
  \citenamefont {Xu}, \citenamefont {Niu},\ and\ \citenamefont
  {Fu}}]{1402-4896-88-6-065008}%
  \BibitemOpen
  \bibfield  {author} {\bibinfo {author} {\bibfnamefont {X~L}\ \bibnamefont
  {Huang}}, \bibinfo {author} {\bibfnamefont {Huan}\ \bibnamefont {Xu}},
  \bibinfo {author} {\bibfnamefont {X~Y}\ \bibnamefont {Niu}}, \ and\ \bibinfo
  {author} {\bibfnamefont {Y~D}\ \bibnamefont {Fu}},\ }\bibfield  {title}
  {\enquote {\bibinfo {title} {A special entangled quantum heat engine based on
  the two-qubit heisenberg xx model},}\ }\href
  {http://stacks.iop.org/1402-4896/88/i=6/a=065008} {\bibfield  {journal}
  {\bibinfo  {journal} {Physica Scripta}\ }\textbf {\bibinfo {volume} {88}},\
  \bibinfo {pages} {065008} (\bibinfo {year} {2013})}\BibitemShut {NoStop}%
\bibitem [{\citenamefont {Thomas}\ and\ \citenamefont
  {Johal}(2011)}]{PhysRevE.83.031135}%
  \BibitemOpen
  \bibfield  {author} {\bibinfo {author} {\bibfnamefont {George}\ \bibnamefont
  {Thomas}}\ and\ \bibinfo {author} {\bibfnamefont {Ramandeep~S.}\ \bibnamefont
  {Johal}},\ }\bibfield  {title} {\enquote {\bibinfo {title} {Coupled quantum
  otto cycle},}\ }\href {\doibase 10.1103/PhysRevE.83.031135} {\bibfield
  {journal} {\bibinfo  {journal} {Phys. Rev. E}\ }\textbf {\bibinfo {volume}
  {83}},\ \bibinfo {pages} {031135} (\bibinfo {year} {2011})}\BibitemShut
  {NoStop}%
\bibitem [{\citenamefont {Scully}\ \emph {et~al.}(2003)\citenamefont {Scully},
  \citenamefont {Zubairy}, \citenamefont {Agarwal},\ and\ \citenamefont
  {Walther}}]{Scully}%
  \BibitemOpen
  \bibfield  {author} {\bibinfo {author} {\bibfnamefont {Marlan~O.}\
  \bibnamefont {Scully}}, \bibinfo {author} {\bibfnamefont {M.~Suhail}\
  \bibnamefont {Zubairy}}, \bibinfo {author} {\bibfnamefont {Girish~S.}\
  \bibnamefont {Agarwal}}, \ and\ \bibinfo {author} {\bibfnamefont {Herbert}\
  \bibnamefont {Walther}},\ }\bibfield  {title} {\enquote {\bibinfo {title}
  {Extracting work from a single heat bath via vanishing quantum coherence},}\
  }\href {\doibase 10.1126/science.1078955} {\bibfield  {journal} {\bibinfo
  {journal} {Science}\ } (\bibinfo {year} {2003}),\
  10.1126/science.1078955}\BibitemShut {NoStop}%
\bibitem [{\citenamefont {Zhang}(2008)}]{Zhang2008}%
  \BibitemOpen
  \bibfield  {author} {\bibinfo {author} {\bibfnamefont {G.~F.}\ \bibnamefont
  {Zhang}},\ }\bibfield  {title} {\enquote {\bibinfo {title} {Entangled quantum
  heat engines based on two two-spin systems with dzyaloshinski-moriya
  anisotropic antisymmetric interaction},}\ }\href {\doibase
  10.1140/epjd/e2008-00133-0} {\bibfield  {journal} {\bibinfo  {journal} {The
  European Physical Journal D}\ }\textbf {\bibinfo {volume} {49}},\ \bibinfo
  {pages} {123} (\bibinfo {year} {2008})}\BibitemShut {NoStop}%
\bibitem [{\citenamefont {{\c{C}}akmak}\ \emph {et~al.}(2016)\citenamefont
  {{\c{C}}akmak}, \citenamefont {Altintas},\ and\ \citenamefont
  {E.~M{\"u}stecapl{\i}o{\u{g}}lu}}]{Cakmak2016}%
  \BibitemOpen
  \bibfield  {author} {\bibinfo {author} {\bibfnamefont {Sel{\c{c}}uk}\
  \bibnamefont {{\c{C}}akmak}}, \bibinfo {author} {\bibfnamefont {Ferdi}\
  \bibnamefont {Altintas}}, \ and\ \bibinfo {author} {\bibfnamefont
  {{\"O}zg{\"u}r}\ \bibnamefont {E.~M{\"u}stecapl{\i}o{\u{g}}lu}},\ }\bibfield
  {title} {\enquote {\bibinfo {title} {Lipkin-meshkov-glick model in a quantum
  otto cycle},}\ }\href {\doibase 10.1140/epjp/i2016-16197-0} {\bibfield
  {journal} {\bibinfo  {journal} {The European Physical Journal Plus}\ }\textbf
  {\bibinfo {volume} {131}},\ \bibinfo {pages} {197} (\bibinfo {year}
  {2016})}\BibitemShut {NoStop}%
\bibitem [{\citenamefont {Zhang}\ \emph {et~al.}(2014)\citenamefont {Zhang},
  \citenamefont {Bariani},\ and\ \citenamefont
  {Meystre}}]{PhysRevLett.112.150602}%
  \BibitemOpen
  \bibfield  {author} {\bibinfo {author} {\bibfnamefont {Keye}\ \bibnamefont
  {Zhang}}, \bibinfo {author} {\bibfnamefont {Francesco}\ \bibnamefont
  {Bariani}}, \ and\ \bibinfo {author} {\bibfnamefont {Pierre}\ \bibnamefont
  {Meystre}},\ }\bibfield  {title} {\enquote {\bibinfo {title} {Quantum
  optomechanical heat engine},}\ }\href {\doibase
  10.1103/PhysRevLett.112.150602} {\bibfield  {journal} {\bibinfo  {journal}
  {Phys. Rev. Lett.}\ }\textbf {\bibinfo {volume} {112}},\ \bibinfo {pages}
  {150602} (\bibinfo {year} {2014})}\BibitemShut {NoStop}%
\bibitem [{\citenamefont {T\"urkpen\ifmmode~\mbox{\c{c}}\else \c{c}\fi{}e}\
  and\ \citenamefont {M\"ustecapl\ifmmode \imath \else \i
  \fi{}o\ifmmode~\breve{g}\else \u{g}\fi{}lu}(2016)}]{PhysRevE.93.012145}%
  \BibitemOpen
  \bibfield  {author} {\bibinfo {author} {\bibfnamefont {Deniz}\ \bibnamefont
  {T\"urkpen\ifmmode~\mbox{\c{c}}\else \c{c}\fi{}e}}\ and\ \bibinfo {author}
  {\bibfnamefont {\"Ozg\"ur~E.}\ \bibnamefont {M\"ustecapl\ifmmode \imath \else
  \i \fi{}o\ifmmode~\breve{g}\else \u{g}\fi{}lu}},\ }\bibfield  {title}
  {\enquote {\bibinfo {title} {Quantum fuel with multilevel atomic coherence
  for ultrahigh specific work in a photonic carnot engine},}\ }\href {\doibase
  10.1103/PhysRevE.93.012145} {\bibfield  {journal} {\bibinfo  {journal} {Phys.
  Rev. E}\ }\textbf {\bibinfo {volume} {93}},\ \bibinfo {pages} {012145}
  (\bibinfo {year} {2016})}\BibitemShut {NoStop}%
\bibitem [{\citenamefont {Chand}\ and\ \citenamefont
  {Biswas}(2017{\natexlab{a}})}]{0295-5075-118-6-60003}%
  \BibitemOpen
  \bibfield  {author} {\bibinfo {author} {\bibfnamefont {Suman}\ \bibnamefont
  {Chand}}\ and\ \bibinfo {author} {\bibfnamefont {Asoka}\ \bibnamefont
  {Biswas}},\ }\bibfield  {title} {\enquote {\bibinfo {title} {Single-ion
  quantum otto engine with always-on bath interaction},}\ }\href
  {http://stacks.iop.org/0295-5075/118/i=6/a=60003} {\bibfield  {journal}
  {\bibinfo  {journal} {EPL (Europhysics Letters)}\ }\textbf {\bibinfo {volume}
  {118}},\ \bibinfo {pages} {60003} (\bibinfo {year}
  {2017}{\natexlab{a}})}\BibitemShut {NoStop}%
\bibitem [{\citenamefont {Chand}\ and\ \citenamefont
  {Biswas}(2017{\natexlab{b}})}]{PhysRevE.95.032111}%
  \BibitemOpen
  \bibfield  {author} {\bibinfo {author} {\bibfnamefont {Suman}\ \bibnamefont
  {Chand}}\ and\ \bibinfo {author} {\bibfnamefont {Asoka}\ \bibnamefont
  {Biswas}},\ }\bibfield  {title} {\enquote {\bibinfo {title}
  {Measurement-induced operation of two-ion quantum heat machines},}\ }\href
  {\doibase 10.1103/PhysRevE.95.032111} {\bibfield  {journal} {\bibinfo
  {journal} {Phys. Rev. E}\ }\textbf {\bibinfo {volume} {95}},\ \bibinfo
  {pages} {032111} (\bibinfo {year} {2017}{\natexlab{b}})}\BibitemShut
  {NoStop}%
\bibitem [{\citenamefont {Uzdin}\ \emph {et~al.}(2015)\citenamefont {Uzdin},
  \citenamefont {Levy},\ and\ \citenamefont {Kosloff}}]{PhysRevX.5.031044}%
  \BibitemOpen
  \bibfield  {author} {\bibinfo {author} {\bibfnamefont {Raam}\ \bibnamefont
  {Uzdin}}, \bibinfo {author} {\bibfnamefont {Amikam}\ \bibnamefont {Levy}}, \
  and\ \bibinfo {author} {\bibfnamefont {Ronnie}\ \bibnamefont {Kosloff}},\
  }\bibfield  {title} {\enquote {\bibinfo {title} {Equivalence of quantum heat
  machines, and quantum-thermodynamic signatures},}\ }\href {\doibase
  10.1103/PhysRevX.5.031044} {\bibfield  {journal} {\bibinfo  {journal} {Phys.
  Rev. X}\ }\textbf {\bibinfo {volume} {5}},\ \bibinfo {pages} {031044}
  (\bibinfo {year} {2015})}\BibitemShut {NoStop}%
\bibitem [{\citenamefont {Alecce}\ \emph {et~al.}(2015)\citenamefont {Alecce},
  \citenamefont {Galve}, \citenamefont {Gullo}, \citenamefont {Dell'Anna},
  \citenamefont {Plastina},\ and\ \citenamefont {Zambrini}}]{Alecce2015}%
  \BibitemOpen
  \bibfield  {author} {\bibinfo {author} {\bibfnamefont {A.}~\bibnamefont
  {Alecce}}, \bibinfo {author} {\bibfnamefont {F.}~\bibnamefont {Galve}},
  \bibinfo {author} {\bibfnamefont {N.~Lo}\ \bibnamefont {Gullo}}, \bibinfo
  {author} {\bibfnamefont {L.}~\bibnamefont {Dell'Anna}}, \bibinfo {author}
  {\bibfnamefont {F.}~\bibnamefont {Plastina}}, \ and\ \bibinfo {author}
  {\bibfnamefont {R.}~\bibnamefont {Zambrini}},\ }\bibfield  {title} {\enquote
  {\bibinfo {title} {{Quantum Otto cycle with inner friction: Finite-time and
  disorder effects}},}\ }\href {\doibase 10.1088/1367-2630/17/7/075007}
  {\bibfield  {journal} {\bibinfo  {journal} {New J. Phys.}\ }\textbf {\bibinfo
  {volume} {17}} (\bibinfo {year} {2015}),\ 10.1088/1367-2630/17/7/075007},\
  \Eprint {http://arxiv.org/abs/1507.03417} {1507.03417} \BibitemShut {NoStop}%
\bibitem [{\citenamefont {Scovil}\ and\ \citenamefont
  {Schulz-DuBois}(1959)}]{PhysRevLett.2.262}%
  \BibitemOpen
  \bibfield  {author} {\bibinfo {author} {\bibfnamefont {H.~E.~D.}\
  \bibnamefont {Scovil}}\ and\ \bibinfo {author} {\bibfnamefont {E.~O.}\
  \bibnamefont {Schulz-DuBois}},\ }\bibfield  {title} {\enquote {\bibinfo
  {title} {Three-level masers as heat engines},}\ }\href {\doibase
  10.1103/PhysRevLett.2.262} {\bibfield  {journal} {\bibinfo  {journal} {Phys.
  Rev. Lett.}\ }\textbf {\bibinfo {volume} {2}},\ \bibinfo {pages} {262--263}
  (\bibinfo {year} {1959})}\BibitemShut {NoStop}%
\bibitem [{\citenamefont {Alicki}(1979)}]{0305-4470-12-5-007}%
  \BibitemOpen
  \bibfield  {author} {\bibinfo {author} {\bibfnamefont {R}~\bibnamefont
  {Alicki}},\ }\bibfield  {title} {\enquote {\bibinfo {title} {The quantum open
  system as a model of the heat engine},}\ }\href
  {http://stacks.iop.org/0305-4470/12/i=5/a=007} {\bibfield  {journal}
  {\bibinfo  {journal} {Journal of Physics A: Mathematical and General}\
  }\textbf {\bibinfo {volume} {12}},\ \bibinfo {pages} {L103} (\bibinfo {year}
  {1979})}\BibitemShut {NoStop}%
\bibitem [{\citenamefont {Linden}\ \emph {et~al.}(2010)\citenamefont {Linden},
  \citenamefont {Popescu},\ and\ \citenamefont
  {Skrzypczyk}}]{PhysRevLett.105.130401}%
  \BibitemOpen
  \bibfield  {author} {\bibinfo {author} {\bibfnamefont {Noah}\ \bibnamefont
  {Linden}}, \bibinfo {author} {\bibfnamefont {Sandu}\ \bibnamefont {Popescu}},
  \ and\ \bibinfo {author} {\bibfnamefont {Paul}\ \bibnamefont {Skrzypczyk}},\
  }\bibfield  {title} {\enquote {\bibinfo {title} {How small can thermal
  machines be? the smallest possible refrigerator},}\ }\href {\doibase
  10.1103/PhysRevLett.105.130401} {\bibfield  {journal} {\bibinfo  {journal}
  {Phys. Rev. Lett.}\ }\textbf {\bibinfo {volume} {105}},\ \bibinfo {pages}
  {130401} (\bibinfo {year} {2010})}\BibitemShut {NoStop}%
\bibitem [{\citenamefont {Uzdin}\ and\ \citenamefont
  {Kosloff}(2014)}]{1367-2630-16-9-095003}%
  \BibitemOpen
  \bibfield  {author} {\bibinfo {author} {\bibfnamefont {Raam}\ \bibnamefont
  {Uzdin}}\ and\ \bibinfo {author} {\bibfnamefont {Ronnie}\ \bibnamefont
  {Kosloff}},\ }\bibfield  {title} {\enquote {\bibinfo {title} {The multilevel
  four-stroke swap engine and its environment},}\ }\href
  {http://stacks.iop.org/1367-2630/16/i=9/a=095003} {\bibfield  {journal}
  {\bibinfo  {journal} {New Journal of Physics}\ }\textbf {\bibinfo {volume}
  {16}},\ \bibinfo {pages} {095003} (\bibinfo {year} {2014})}\BibitemShut
  {NoStop}%
\bibitem [{\citenamefont {Gelbwaser-Klimovsky}\ \emph
  {et~al.}(2013)\citenamefont {Gelbwaser-Klimovsky}, \citenamefont {Alicki},\
  and\ \citenamefont {Kurizki}}]{PhysRevE.87.012140}%
  \BibitemOpen
  \bibfield  {author} {\bibinfo {author} {\bibfnamefont {D.}~\bibnamefont
  {Gelbwaser-Klimovsky}}, \bibinfo {author} {\bibfnamefont {R.}~\bibnamefont
  {Alicki}}, \ and\ \bibinfo {author} {\bibfnamefont {G.}~\bibnamefont
  {Kurizki}},\ }\bibfield  {title} {\enquote {\bibinfo {title} {Minimal
  universal quantum heat machine},}\ }\href {\doibase
  10.1103/PhysRevE.87.012140} {\bibfield  {journal} {\bibinfo  {journal} {Phys.
  Rev. E}\ }\textbf {\bibinfo {volume} {87}},\ \bibinfo {pages} {012140}
  (\bibinfo {year} {2013})}\BibitemShut {NoStop}%
\bibitem [{\citenamefont {Kosloff}\ and\ \citenamefont
  {Levy}(2014)}]{doi:10.1146/annurev-physchem-040513-103724}%
  \BibitemOpen
  \bibfield  {author} {\bibinfo {author} {\bibfnamefont {Ronnie}\ \bibnamefont
  {Kosloff}}\ and\ \bibinfo {author} {\bibfnamefont {Amikam}\ \bibnamefont
  {Levy}},\ }\bibfield  {title} {\enquote {\bibinfo {title} {Quantum heat
  engines and refrigerators: Continuous devices},}\ }\href {\doibase
  10.1146/annurev-physchem-040513-103724} {\bibfield  {journal} {\bibinfo
  {journal} {Annual Review of Physical Chemistry}\ }\textbf {\bibinfo {volume}
  {65}},\ \bibinfo {pages} {365--393} (\bibinfo {year} {2014})},\ \bibinfo
  {note} {pMID: 24689798}\BibitemShut {NoStop}%
\bibitem [{\citenamefont {Feldmann}\ and\ \citenamefont
  {Kosloff}(2003)}]{PhysRevE.68.016101}%
  \BibitemOpen
  \bibfield  {author} {\bibinfo {author} {\bibfnamefont {Tova}\ \bibnamefont
  {Feldmann}}\ and\ \bibinfo {author} {\bibfnamefont {Ronnie}\ \bibnamefont
  {Kosloff}},\ }\bibfield  {title} {\enquote {\bibinfo {title} {Quantum
  four-stroke heat engine: Thermodynamic observables in a model with intrinsic
  friction},}\ }\href {\doibase 10.1103/PhysRevE.68.016101} {\bibfield
  {journal} {\bibinfo  {journal} {Phys. Rev. E}\ }\textbf {\bibinfo {volume}
  {68}},\ \bibinfo {pages} {016101} (\bibinfo {year} {2003})}\BibitemShut
  {NoStop}%
\bibitem [{\citenamefont {Ro{\ss}nagel}\ \emph {et~al.}(2016)\citenamefont
  {Ro{\ss}nagel}, \citenamefont {Dawkins}, \citenamefont {Tolazzi},
  \citenamefont {Abah}, \citenamefont {Lutz}, \citenamefont {Schmidt-Kaler},\
  and\ \citenamefont {Singer}}]{Ronagel325}%
  \BibitemOpen
  \bibfield  {author} {\bibinfo {author} {\bibfnamefont {Johannes}\
  \bibnamefont {Ro{\ss}nagel}}, \bibinfo {author} {\bibfnamefont {Samuel~T.}\
  \bibnamefont {Dawkins}}, \bibinfo {author} {\bibfnamefont {Karl~N.}\
  \bibnamefont {Tolazzi}}, \bibinfo {author} {\bibfnamefont {Obinna}\
  \bibnamefont {Abah}}, \bibinfo {author} {\bibfnamefont {Eric}\ \bibnamefont
  {Lutz}}, \bibinfo {author} {\bibfnamefont {Ferdinand}\ \bibnamefont
  {Schmidt-Kaler}}, \ and\ \bibinfo {author} {\bibfnamefont {Kilian}\
  \bibnamefont {Singer}},\ }\bibfield  {title} {\enquote {\bibinfo {title} {A
  single-atom heat engine},}\ }\href {\doibase 10.1126/science.aad6320}
  {\bibfield  {journal} {\bibinfo  {journal} {Science}\ }\textbf {\bibinfo
  {volume} {352}},\ \bibinfo {pages} {325--329} (\bibinfo {year} {2016})},\
  \Eprint
  {http://arxiv.org/abs/http://science.sciencemag.org/content/352/6283/325.full.pdf}
  {http://science.sciencemag.org/content/352/6283/325.full.pdf} \BibitemShut
  {NoStop}%
\bibitem [{\citenamefont {Abah}\ \emph {et~al.}(2012)\citenamefont {Abah},
  \citenamefont {Ro\ss{}nagel}, \citenamefont {Jacob}, \citenamefont {Deffner},
  \citenamefont {Schmidt-Kaler}, \citenamefont {Singer},\ and\ \citenamefont
  {Lutz}}]{PhysRevLett.109.203006}%
  \BibitemOpen
  \bibfield  {author} {\bibinfo {author} {\bibfnamefont {O.}~\bibnamefont
  {Abah}}, \bibinfo {author} {\bibfnamefont {J.}~\bibnamefont {Ro\ss{}nagel}},
  \bibinfo {author} {\bibfnamefont {G.}~\bibnamefont {Jacob}}, \bibinfo
  {author} {\bibfnamefont {S.}~\bibnamefont {Deffner}}, \bibinfo {author}
  {\bibfnamefont {F.}~\bibnamefont {Schmidt-Kaler}}, \bibinfo {author}
  {\bibfnamefont {K.}~\bibnamefont {Singer}}, \ and\ \bibinfo {author}
  {\bibfnamefont {E.}~\bibnamefont {Lutz}},\ }\bibfield  {title} {\enquote
  {\bibinfo {title} {Single-ion heat engine at maximum power},}\ }\href
  {\doibase 10.1103/PhysRevLett.109.203006} {\bibfield  {journal} {\bibinfo
  {journal} {Phys. Rev. Lett.}\ }\textbf {\bibinfo {volume} {109}},\ \bibinfo
  {pages} {203006} (\bibinfo {year} {2012})}\BibitemShut {NoStop}%
\bibitem [{\citenamefont {Klaers}\ \emph {et~al.}(2017)\citenamefont {Klaers},
  \citenamefont {Faelt}, \citenamefont {Imamoglu},\ and\ \citenamefont
  {Togan}}]{PhysRevX.7.031044}%
  \BibitemOpen
  \bibfield  {author} {\bibinfo {author} {\bibfnamefont {Jan}\ \bibnamefont
  {Klaers}}, \bibinfo {author} {\bibfnamefont {Stefan}\ \bibnamefont {Faelt}},
  \bibinfo {author} {\bibfnamefont {Atac}\ \bibnamefont {Imamoglu}}, \ and\
  \bibinfo {author} {\bibfnamefont {Emre}\ \bibnamefont {Togan}},\ }\bibfield
  {title} {\enquote {\bibinfo {title} {Squeezed thermal reservoirs as a
  resource for a nanomechanical engine beyond the carnot limit},}\ }\href
  {\doibase 10.1103/PhysRevX.7.031044} {\bibfield  {journal} {\bibinfo
  {journal} {Phys. Rev. X}\ }\textbf {\bibinfo {volume} {7}},\ \bibinfo {pages}
  {031044} (\bibinfo {year} {2017})}\BibitemShut {NoStop}%
\bibitem [{\citenamefont {Ro\ss{}nagel}\ \emph {et~al.}(2014)\citenamefont
  {Ro\ss{}nagel}, \citenamefont {Abah}, \citenamefont {Schmidt-Kaler},
  \citenamefont {Singer},\ and\ \citenamefont {Lutz}}]{PhysRevLett.112.030602}%
  \BibitemOpen
  \bibfield  {author} {\bibinfo {author} {\bibfnamefont {J.}~\bibnamefont
  {Ro\ss{}nagel}}, \bibinfo {author} {\bibfnamefont {O.}~\bibnamefont {Abah}},
  \bibinfo {author} {\bibfnamefont {F.}~\bibnamefont {Schmidt-Kaler}}, \bibinfo
  {author} {\bibfnamefont {K.}~\bibnamefont {Singer}}, \ and\ \bibinfo {author}
  {\bibfnamefont {E.}~\bibnamefont {Lutz}},\ }\bibfield  {title} {\enquote
  {\bibinfo {title} {Nanoscale heat engine beyond the carnot limit},}\ }\href
  {\doibase 10.1103/PhysRevLett.112.030602} {\bibfield  {journal} {\bibinfo
  {journal} {Phys. Rev. Lett.}\ }\textbf {\bibinfo {volume} {112}},\ \bibinfo
  {pages} {030602} (\bibinfo {year} {2014})}\BibitemShut {NoStop}%
\bibitem [{\citenamefont {Dillenschneider}\ and\ \citenamefont
  {Lutz}(2009)}]{0295-5075-88-5-50003}%
  \BibitemOpen
  \bibfield  {author} {\bibinfo {author} {\bibfnamefont {R.}~\bibnamefont
  {Dillenschneider}}\ and\ \bibinfo {author} {\bibfnamefont {E.}~\bibnamefont
  {Lutz}},\ }\bibfield  {title} {\enquote {\bibinfo {title} {Energetics of
  quantum correlations},}\ }\href
  {http://stacks.iop.org/0295-5075/88/i=5/a=50003} {\bibfield  {journal}
  {\bibinfo  {journal} {EPL (Europhysics Letters)}\ }\textbf {\bibinfo {volume}
  {88}},\ \bibinfo {pages} {50003} (\bibinfo {year} {2009})}\BibitemShut
  {NoStop}%
\bibitem [{\citenamefont {Dağ}\ \emph {et~al.}(2016)\citenamefont {Dağ},
  \citenamefont {Niedenzu}, \citenamefont {M{\"{u}}stecaplioğlu},\ and\
  \citenamefont {Kurizki}}]{Dag2016}%
  \BibitemOpen
  \bibfield  {author} {\bibinfo {author} {\bibfnamefont {Ceren~B.}\
  \bibnamefont {Dağ}}, \bibinfo {author} {\bibfnamefont {Wolfgang}\
  \bibnamefont {Niedenzu}}, \bibinfo {author} {\bibfnamefont
  {{\"{O}}zg{\"{u}}r~E.}\ \bibnamefont {M{\"{u}}stecaplioğlu}}, \ and\
  \bibinfo {author} {\bibfnamefont {Gershon}\ \bibnamefont {Kurizki}},\
  }\bibfield  {title} {\enquote {\bibinfo {title} {{Multiatom quantum
  coherences in micromasers as fuel for thermal and nonthermal machines}},}\
  }\href {\doibase 10.3390/e18070244} {\bibfield  {journal} {\bibinfo
  {journal} {Entropy}\ }\textbf {\bibinfo {volume} {18}},\ \bibinfo {pages}
  {244} (\bibinfo {year} {2016})}\BibitemShut {NoStop}%
\bibitem [{\citenamefont {Peterson}\ \emph {et~al.}(2018)\citenamefont
  {Peterson}, \citenamefont {Batalh{\~{a}}o}, \citenamefont {Herrera},
  \citenamefont {Souza}, \citenamefont {Sarthour}, \citenamefont {Oliveira},\
  and\ \citenamefont {Serra}}]{Peterson2018}%
  \BibitemOpen
  \bibfield  {author} {\bibinfo {author} {\bibfnamefont {John P.~S.}\
  \bibnamefont {Peterson}}, \bibinfo {author} {\bibfnamefont {Tiago~B.}\
  \bibnamefont {Batalh{\~{a}}o}}, \bibinfo {author} {\bibfnamefont {Marcela}\
  \bibnamefont {Herrera}}, \bibinfo {author} {\bibfnamefont {Alexandre~M.}\
  \bibnamefont {Souza}}, \bibinfo {author} {\bibfnamefont {Roberto~S.}\
  \bibnamefont {Sarthour}}, \bibinfo {author} {\bibfnamefont {Ivan~S.}\
  \bibnamefont {Oliveira}}, \ and\ \bibinfo {author} {\bibfnamefont
  {Roberto~M.}\ \bibnamefont {Serra}},\ }\bibfield  {title} {\enquote {\bibinfo
  {title} {{Experimental characterization of a spin quantum heat engine}},}\
  }\href {https://arxiv.org/abs/1803.06021} {\  (\bibinfo {year} {2018})},\
  \Eprint {http://arxiv.org/abs/1803.06021} {arXiv:1803.06021} \BibitemShut
  {NoStop}%
\bibitem [{\citenamefont {{De Assis}}\ \emph {et~al.}(2018)\citenamefont {{De
  Assis}}, \citenamefont {{De Mendon{\c{c}}a}}, \citenamefont {Villas-Boas},
  \citenamefont {{De Souza}}, \citenamefont {Sarthour}, \citenamefont
  {Oliveira},\ and\ \citenamefont {{De Almeida}}}]{DeAssis2018}%
  \BibitemOpen
  \bibfield  {author} {\bibinfo {author} {\bibfnamefont {Rog{\'{e}}rio~J}\
  \bibnamefont {{De Assis}}}, \bibinfo {author} {\bibfnamefont {Taysa~M}\
  \bibnamefont {{De Mendon{\c{c}}a}}}, \bibinfo {author} {\bibfnamefont
  {Celso~J}\ \bibnamefont {Villas-Boas}}, \bibinfo {author} {\bibfnamefont
  {Alexandre~M}\ \bibnamefont {{De Souza}}}, \bibinfo {author} {\bibfnamefont
  {Roberto~S}\ \bibnamefont {Sarthour}}, \bibinfo {author} {\bibfnamefont
  {Ivan~S}\ \bibnamefont {Oliveira}}, \ and\ \bibinfo {author} {\bibfnamefont
  {Norton~G}\ \bibnamefont {{De Almeida}}},\ }\bibfield  {title} {\enquote
  {\bibinfo {title} {{Quantum heating engine beating the Otto limit}},}\ }\href
  {https://arxiv.org/pdf/1811.02917.pdf} {\  (\bibinfo {year} {2018})},\
  \Eprint {http://arxiv.org/abs/1811.02917v1} {arXiv:1811.02917v1} \BibitemShut
  {NoStop}%
\bibitem [{\citenamefont {{\c{C}}akmak}\ \emph {et~al.}(2017)\citenamefont
  {{\c{C}}akmak}, \citenamefont {Altintas}, \citenamefont {Gen{\c{c}}ten},\
  and\ \citenamefont {M{\"u}stecapl{\i}o{\u{g}}lu}}]{Cakmak2017}%
  \BibitemOpen
  \bibfield  {author} {\bibinfo {author} {\bibfnamefont {Sel{\c{c}}uk}\
  \bibnamefont {{\c{C}}akmak}}, \bibinfo {author} {\bibfnamefont {Ferdi}\
  \bibnamefont {Altintas}}, \bibinfo {author} {\bibfnamefont {Azmi}\
  \bibnamefont {Gen{\c{c}}ten}}, \ and\ \bibinfo {author} {\bibfnamefont
  {{\"O}zg{\"u}r~E.}\ \bibnamefont {M{\"u}stecapl{\i}o{\u{g}}lu}},\ }\bibfield
  {title} {\enquote {\bibinfo {title} {Irreversible work and internal friction
  in a quantum otto cycle of a single arbitrary spin},}\ }\href {\doibase
  10.1140/epjd/e2017-70443-1} {\bibfield  {journal} {\bibinfo  {journal} {The
  European Physical Journal D}\ }\textbf {\bibinfo {volume} {71}},\ \bibinfo
  {pages} {75} (\bibinfo {year} {2017})}\BibitemShut {NoStop}%
\bibitem [{\citenamefont {Park}\ \emph {et~al.}(2016)\citenamefont {Park},
  \citenamefont {Rodriguez-Briones}, \citenamefont {Feng}, \citenamefont
  {Rahimi}, \citenamefont {Baugh},\ and\ \citenamefont {Laflamme}}]{Park2016}%
  \BibitemOpen
  \bibfield  {author} {\bibinfo {author} {\bibfnamefont {Daniel~K.}\
  \bibnamefont {Park}}, \bibinfo {author} {\bibfnamefont {Nayeli~A.}\
  \bibnamefont {Rodriguez-Briones}}, \bibinfo {author} {\bibfnamefont {Guanru}\
  \bibnamefont {Feng}}, \bibinfo {author} {\bibfnamefont {Robabeh}\
  \bibnamefont {Rahimi}}, \bibinfo {author} {\bibfnamefont {Jonathan}\
  \bibnamefont {Baugh}}, \ and\ \bibinfo {author} {\bibfnamefont {Raymond}\
  \bibnamefont {Laflamme}},\ }\enquote {\bibinfo {title} {Heat bath algorithmic
  cooling with spins: Review and prospects},}\ in\ \href {\doibase
  10.1007/978-1-4939-3658-8_8} {\emph {\bibinfo {booktitle} {Electron Spin
  Resonance (ESR) Based Quantum Computing}}},\ \bibinfo {editor} {edited by\
  \bibinfo {editor} {\bibfnamefont {Takeji}\ \bibnamefont {Takui}}, \bibinfo
  {editor} {\bibfnamefont {Lawrence}\ \bibnamefont {Berliner}}, \ and\ \bibinfo
  {editor} {\bibfnamefont {Graeme}\ \bibnamefont {Hanson}}}\ (\bibinfo
  {publisher} {Springer New York},\ \bibinfo {address} {New York, NY},\
  \bibinfo {year} {2016})\ pp.\ \bibinfo {pages} {227--255}\BibitemShut
  {NoStop}%
\bibitem [{\citenamefont {Rodr\'{\i}guez-Briones}\ and\ \citenamefont
  {Laflamme}(2016)}]{PhysRevLett.116.170501}%
  \BibitemOpen
  \bibfield  {author} {\bibinfo {author} {\bibfnamefont {Nayeli~Azucena}\
  \bibnamefont {Rodr\'{\i}guez-Briones}}\ and\ \bibinfo {author} {\bibfnamefont
  {Raymond}\ \bibnamefont {Laflamme}},\ }\bibfield  {title} {\enquote {\bibinfo
  {title} {Achievable polarization for heat-bath algorithmic cooling},}\ }\href
  {\doibase 10.1103/PhysRevLett.116.170501} {\bibfield  {journal} {\bibinfo
  {journal} {Phys. Rev. Lett.}\ }\textbf {\bibinfo {volume} {116}},\ \bibinfo
  {pages} {170501} (\bibinfo {year} {2016})}\BibitemShut {NoStop}%
\bibitem [{\citenamefont {Atia}\ \emph {et~al.}(2016)\citenamefont {Atia},
  \citenamefont {Elias}, \citenamefont {Mor},\ and\ \citenamefont
  {Weinstein}}]{PhysRevA.93.012325}%
  \BibitemOpen
  \bibfield  {author} {\bibinfo {author} {\bibfnamefont {Yosi}\ \bibnamefont
  {Atia}}, \bibinfo {author} {\bibfnamefont {Yuval}\ \bibnamefont {Elias}},
  \bibinfo {author} {\bibfnamefont {Tal}\ \bibnamefont {Mor}}, \ and\ \bibinfo
  {author} {\bibfnamefont {Yossi}\ \bibnamefont {Weinstein}},\ }\bibfield
  {title} {\enquote {\bibinfo {title} {Algorithmic cooling in liquid-state
  nuclear magnetic resonance},}\ }\href {\doibase 10.1103/PhysRevA.93.012325}
  {\bibfield  {journal} {\bibinfo  {journal} {Phys. Rev. A}\ }\textbf {\bibinfo
  {volume} {93}},\ \bibinfo {pages} {012325} (\bibinfo {year}
  {2016})}\BibitemShut {NoStop}%
\bibitem [{\citenamefont {Brassard}\ \emph
  {et~al.}(2014{\natexlab{a}})\citenamefont {Brassard}, \citenamefont {Elias},
  \citenamefont {Fernandez}, \citenamefont {Gilboa}, \citenamefont {Jones},
  \citenamefont {Mor}, \citenamefont {Weinstein},\ and\ \citenamefont
  {Xiao}}]{Brassard20144}%
  \BibitemOpen
  \bibfield  {author} {\bibinfo {author} {\bibfnamefont {G.}~\bibnamefont
  {Brassard}}, \bibinfo {author} {\bibfnamefont {Y.}~\bibnamefont {Elias}},
  \bibinfo {author} {\bibfnamefont {J.~M.}\ \bibnamefont {Fernandez}}, \bibinfo
  {author} {\bibfnamefont {H.}~\bibnamefont {Gilboa}}, \bibinfo {author}
  {\bibfnamefont {J.~A.}\ \bibnamefont {Jones}}, \bibinfo {author}
  {\bibfnamefont {T.}~\bibnamefont {Mor}}, \bibinfo {author} {\bibfnamefont
  {Y.}~\bibnamefont {Weinstein}}, \ and\ \bibinfo {author} {\bibfnamefont
  {L.}~\bibnamefont {Xiao}},\ }\bibfield  {title} {\enquote {\bibinfo {title}
  {Experimental heat-bath cooling of spins},}\ }\href {\doibase
  10.1140/epjp/i2014-14266-0} {\bibfield  {journal} {\bibinfo  {journal} {The
  European Physical Journal Plus}\ }\textbf {\bibinfo {volume} {129}},\
  \bibinfo {pages} {266} (\bibinfo {year} {2014}{\natexlab{a}})}\BibitemShut
  {NoStop}%
\bibitem [{\citenamefont {Brassard}\ \emph
  {et~al.}(2014{\natexlab{b}})\citenamefont {Brassard}, \citenamefont {Elias},
  \citenamefont {Mor},\ and\ \citenamefont {Weinstein}}]{Brassard2014}%
  \BibitemOpen
  \bibfield  {author} {\bibinfo {author} {\bibfnamefont {Gilles}\ \bibnamefont
  {Brassard}}, \bibinfo {author} {\bibfnamefont {Yuval}\ \bibnamefont {Elias}},
  \bibinfo {author} {\bibfnamefont {Tal}\ \bibnamefont {Mor}}, \ and\ \bibinfo
  {author} {\bibfnamefont {Yossi}\ \bibnamefont {Weinstein}},\ }\bibfield
  {title} {\enquote {\bibinfo {title} {Prospects and limitations of algorithmic
  cooling},}\ }\href {\doibase 10.1140/epjp/i2014-14258-0} {\bibfield
  {journal} {\bibinfo  {journal} {The European Physical Journal Plus}\ }\textbf
  {\bibinfo {volume} {129}},\ \bibinfo {pages} {258} (\bibinfo {year}
  {2014}{\natexlab{b}})}\BibitemShut {NoStop}%
\bibitem [{\citenamefont {Fernandez}\ \emph {et~al.}(2004)\citenamefont
  {Fernandez}, \citenamefont {Lloyd}, \citenamefont {Mor},\ and\ \citenamefont
  {Roychowdhury}}]{Fernandez2004}%
  \BibitemOpen
  \bibfield  {author} {\bibinfo {author} {\bibfnamefont {Jose~M.}\ \bibnamefont
  {Fernandez}}, \bibinfo {author} {\bibfnamefont {Seth}\ \bibnamefont {Lloyd}},
  \bibinfo {author} {\bibfnamefont {Tal}\ \bibnamefont {Mor}}, \ and\ \bibinfo
  {author} {\bibfnamefont {Vwani}\ \bibnamefont {Roychowdhury}},\ }\bibfield
  {title} {\enquote {\bibinfo {title} {{Algorithmic Cooling of Spins: A
  Practicable Method for Increasing Polarization}},}\ }\href {\doibase
  10.1142/S0219749904000419} {\bibfield  {journal} {\bibinfo  {journal} {Int.
  J. Quantum Inf.}\ }\textbf {\bibinfo {volume} {02}},\ \bibinfo {pages}
  {461--477} (\bibinfo {year} {2004})}\BibitemShut {NoStop}%
\bibitem [{\citenamefont {Elias}\ \emph {et~al.}(2007)\citenamefont {Elias},
  \citenamefont {Fernandez}, \citenamefont {Mor},\ and\ \citenamefont
  {Weinstein}}]{Elias}%
  \BibitemOpen
  \bibfield  {author} {\bibinfo {author} {\bibfnamefont {Yuval}\ \bibnamefont
  {Elias}}, \bibinfo {author} {\bibfnamefont {Jos{\'{e}}~M.}\ \bibnamefont
  {Fernandez}}, \bibinfo {author} {\bibfnamefont {Tal}\ \bibnamefont {Mor}}, \
  and\ \bibinfo {author} {\bibfnamefont {Yossi}\ \bibnamefont {Weinstein}},\
  }\href {\doibase 10.1007/978-3-540-73554-0} {\emph {\bibinfo {title} {Unconv.
  Comput.}}},\ edited by\ \bibinfo {editor} {\bibfnamefont {Selim~G.}\
  \bibnamefont {Akl}}, \bibinfo {editor} {\bibfnamefont {Cristian~S.}\
  \bibnamefont {Calude}}, \bibinfo {editor} {\bibfnamefont {Michael~J.}\
  \bibnamefont {Dinneen}}, \bibinfo {editor} {\bibfnamefont {Grzegorz}\
  \bibnamefont {Rozenberg}}, \ and\ \bibinfo {editor} {\bibfnamefont {H.~Todd}\
  \bibnamefont {Wareham}},\ \bibinfo {series} {Lecture Notes in Computer
  Science}, Vol.\ \bibinfo {volume} {4618}\ (\bibinfo  {publisher} {Springer
  Berlin Heidelberg},\ \bibinfo {address} {Berlin, Heidelberg},\ \bibinfo
  {year} {2007})\ pp.\ \bibinfo {pages} {2--26},\ \Eprint
  {http://arxiv.org/abs/0711.2964} {0711.2964} \BibitemShut {NoStop}%
\bibitem [{\citenamefont {Boykin}\ \emph {et~al.}(2002)\citenamefont {Boykin},
  \citenamefont {Mor}, \citenamefont {Roychowdhury}, \citenamefont {Vatan},\
  and\ \citenamefont {Vrijen}}]{Boykin3388}%
  \BibitemOpen
  \bibfield  {author} {\bibinfo {author} {\bibfnamefont {P.~Oscar}\
  \bibnamefont {Boykin}}, \bibinfo {author} {\bibfnamefont {Tal}\ \bibnamefont
  {Mor}}, \bibinfo {author} {\bibfnamefont {Vwani}\ \bibnamefont
  {Roychowdhury}}, \bibinfo {author} {\bibfnamefont {Farrokh}\ \bibnamefont
  {Vatan}}, \ and\ \bibinfo {author} {\bibfnamefont {Rutger}\ \bibnamefont
  {Vrijen}},\ }\bibfield  {title} {\enquote {\bibinfo {title} {Algorithmic
  cooling and scalable nmr quantum computers},}\ }\href {\doibase
  10.1073/pnas.241641898} {\bibfield  {journal} {\bibinfo  {journal}
  {Proceedings of the National Academy of Sciences}\ }\textbf {\bibinfo
  {volume} {99}},\ \bibinfo {pages} {3388--3393} (\bibinfo {year} {2002})},\
  \Eprint {http://arxiv.org/abs/http://www.pnas.org/content/99/6/3388.full.pdf}
  {http://www.pnas.org/content/99/6/3388.full.pdf} \BibitemShut {NoStop}%
\bibitem [{\citenamefont {Kafri}\ and\ \citenamefont
  {Taylor}(2012)}]{Kafri2012}%
  \BibitemOpen
  \bibfield  {author} {\bibinfo {author} {\bibfnamefont {Dvir}\ \bibnamefont
  {Kafri}}\ and\ \bibinfo {author} {\bibfnamefont {Jacob~M.}\ \bibnamefont
  {Taylor}},\ }\bibfield  {title} {\enquote {\bibinfo {title} {{Algorithmic
  Cooling of a Quantum Simulator}},}\ }\href@noop {} {\ ,\ \bibinfo {pages}
  {41} (\bibinfo {year} {2012})},\ \Eprint {http://arxiv.org/abs/1207.7111}
  {1207.7111} \BibitemShut {NoStop}%
\bibitem [{\citenamefont {Raeisi}\ and\ \citenamefont
  {Mosca}(2015)}]{PhysRevLett.114.100404}%
  \BibitemOpen
  \bibfield  {author} {\bibinfo {author} {\bibfnamefont {Sadegh}\ \bibnamefont
  {Raeisi}}\ and\ \bibinfo {author} {\bibfnamefont {Michele}\ \bibnamefont
  {Mosca}},\ }\bibfield  {title} {\enquote {\bibinfo {title} {Asymptotic bound
  for heat-bath algorithmic cooling},}\ }\href {\doibase
  10.1103/PhysRevLett.114.100404} {\bibfield  {journal} {\bibinfo  {journal}
  {Phys. Rev. Lett.}\ }\textbf {\bibinfo {volume} {114}},\ \bibinfo {pages}
  {100404} (\bibinfo {year} {2015})}\BibitemShut {NoStop}%
\bibitem [{\citenamefont {Rodr{\'{i}}guez-Briones}\ \emph
  {et~al.}(2017)\citenamefont {Rodr{\'{i}}guez-Briones}, \citenamefont {Li},
  \citenamefont {Peng}, \citenamefont {Mor}, \citenamefont {Weinstein},\ and\
  \citenamefont {Laflamme}}]{Rodriguez-Briones2017b}%
  \BibitemOpen
  \bibfield  {author} {\bibinfo {author} {\bibfnamefont {Nayeli~A.}\
  \bibnamefont {Rodr{\'{i}}guez-Briones}}, \bibinfo {author} {\bibfnamefont
  {Jun}\ \bibnamefont {Li}}, \bibinfo {author} {\bibfnamefont {Xinhua}\
  \bibnamefont {Peng}}, \bibinfo {author} {\bibfnamefont {Tal}\ \bibnamefont
  {Mor}}, \bibinfo {author} {\bibfnamefont {Yossi}\ \bibnamefont {Weinstein}},
  \ and\ \bibinfo {author} {\bibfnamefont {Raymond}\ \bibnamefont {Laflamme}},\
  }\bibfield  {title} {\enquote {\bibinfo {title} {{Heat-bath algorithmic
  cooling with correlated qubit-environment interactions}},}\ }\href {\doibase
  10.1088/1367-2630/aa8fe0} {\bibfield  {journal} {\bibinfo  {journal} {New J.
  Phys.}\ }\textbf {\bibinfo {volume} {19}} (\bibinfo {year} {2017}),\
  10.1088/1367-2630/aa8fe0}\BibitemShut {NoStop}%
\bibitem [{\citenamefont {Ryan}\ \emph {et~al.}(2008)\citenamefont {Ryan},
  \citenamefont {Moussa}, \citenamefont {Baugh},\ and\ \citenamefont
  {Laflamme}}]{PhysRevLett.100.140501}%
  \BibitemOpen
  \bibfield  {author} {\bibinfo {author} {\bibfnamefont {C.~A.}\ \bibnamefont
  {Ryan}}, \bibinfo {author} {\bibfnamefont {O.}~\bibnamefont {Moussa}},
  \bibinfo {author} {\bibfnamefont {J.}~\bibnamefont {Baugh}}, \ and\ \bibinfo
  {author} {\bibfnamefont {R.}~\bibnamefont {Laflamme}},\ }\bibfield  {title}
  {\enquote {\bibinfo {title} {Spin based heat engine: Demonstration of
  multiple rounds of algorithmic cooling},}\ }\href {\doibase
  10.1103/PhysRevLett.100.140501} {\bibfield  {journal} {\bibinfo  {journal}
  {Phys. Rev. Lett.}\ }\textbf {\bibinfo {volume} {100}},\ \bibinfo {pages}
  {140501} (\bibinfo {year} {2008})}\BibitemShut {NoStop}%
\bibitem [{\citenamefont {Schulman}\ \emph {et~al.}(2005)\citenamefont
  {Schulman}, \citenamefont {Mor},\ and\ \citenamefont
  {Weinstein}}]{PhysRevLett.94.120501}%
  \BibitemOpen
  \bibfield  {author} {\bibinfo {author} {\bibfnamefont {Leonard~J.}\
  \bibnamefont {Schulman}}, \bibinfo {author} {\bibfnamefont {Tal}\
  \bibnamefont {Mor}}, \ and\ \bibinfo {author} {\bibfnamefont {Yossi}\
  \bibnamefont {Weinstein}},\ }\bibfield  {title} {\enquote {\bibinfo {title}
  {Physical limits of heat-bath algorithmic cooling},}\ }\href {\doibase
  10.1103/PhysRevLett.94.120501} {\bibfield  {journal} {\bibinfo  {journal}
  {Phys. Rev. Lett.}\ }\textbf {\bibinfo {volume} {94}},\ \bibinfo {pages}
  {120501} (\bibinfo {year} {2005})}\BibitemShut {NoStop}%
\bibitem [{\citenamefont {Elias}\ \emph {et~al.}(2006)\citenamefont {Elias},
  \citenamefont {Fernandez}, \citenamefont {Mor},\ and\ \citenamefont
  {Weinstein}}]{Elias2006}%
  \BibitemOpen
  \bibfield  {author} {\bibinfo {author} {\bibfnamefont {Yuval}\ \bibnamefont
  {Elias}}, \bibinfo {author} {\bibfnamefont {Jos{\'{e}}~M.}\ \bibnamefont
  {Fernandez}}, \bibinfo {author} {\bibfnamefont {Tal}\ \bibnamefont {Mor}}, \
  and\ \bibinfo {author} {\bibfnamefont {Yossi}\ \bibnamefont {Weinstein}},\
  }\bibfield  {title} {\enquote {\bibinfo {title} {{Optimal Algorithmic Cooling
  of Spins}},}\ }\href {\doibase 10.1560/IJC_46_4_371} {\bibfield  {journal}
  {\bibinfo  {journal} {Isr. J. Chem.}\ }\textbf {\bibinfo {volume} {46}},\
  \bibinfo {pages} {371--391} (\bibinfo {year} {2006})}\BibitemShut {NoStop}%
\bibitem [{\citenamefont {Oliveira}\ \emph {et~al.}(2007)\citenamefont
  {Oliveira}, \citenamefont {Bonagamba}, \citenamefont {Sarthour},
  \citenamefont {Freitas},\ and\ \citenamefont {deAzevedo}}]{OLIVEIRA2007137}%
  \BibitemOpen
  \bibfield  {author} {\bibinfo {author} {\bibfnamefont {Ivan~S.}\ \bibnamefont
  {Oliveira}}, \bibinfo {author} {\bibfnamefont {Tito~J.}\ \bibnamefont
  {Bonagamba}}, \bibinfo {author} {\bibfnamefont {Roberto~S.}\ \bibnamefont
  {Sarthour}}, \bibinfo {author} {\bibfnamefont {Jair~C.C.}\ \bibnamefont
  {Freitas}}, \ and\ \bibinfo {author} {\bibfnamefont {Eduardo~R.}\
  \bibnamefont {deAzevedo}},\ }\bibfield  {title} {\enquote {\bibinfo {title}
  {4 - introduction to nmr quantum computing},}\ }in\ \href {\doibase
  https://doi.org/10.1016/B978-044452782-0/50006-3} {\emph {\bibinfo
  {booktitle} {NMR Quantum Information Processing}}}\ (\bibinfo  {publisher}
  {Elsevier Science B.V.},\ \bibinfo {address} {Amsterdam},\ \bibinfo {year}
  {2007})\ pp.\ \bibinfo {pages} {137 -- 181}\BibitemShut {NoStop}%
\bibitem [{\citenamefont {Devoret}\ \emph {et~al.}(2011)\citenamefont
  {Devoret}, \citenamefont {Huard}, \citenamefont {Schoelkopf},\ and\
  \citenamefont {Cugliandolo}}]{Devoret2011}%
  \BibitemOpen
  \bibfield  {author} {\bibinfo {author} {\bibfnamefont {M.}~\bibnamefont
  {Devoret}}, \bibinfo {author} {\bibfnamefont {B.}~\bibnamefont {Huard}},
  \bibinfo {author} {\bibfnamefont {R.}~\bibnamefont {Schoelkopf}}, \ and\
  \bibinfo {author} {\bibfnamefont {L.F.}\ \bibnamefont {Cugliandolo}},\ }\href
  {\doibase 10.1093/acprof:oso/9780199681181.001.0001} {\emph {\bibinfo {title}
  {Quantum Machines: Measurement and Control of Engineered Quantum Systems}}}\
  (\bibinfo {year} {2011})\BibitemShut {NoStop}%
\bibitem [{\citenamefont {Abragam}(1963)}]{Abragam1963}%
  \BibitemOpen
  \bibfield  {author} {\bibinfo {author} {\bibfnamefont {A}~\bibnamefont
  {Abragam}},\ }\bibfield  {title} {\enquote {\bibinfo {title} {{Principles of
  nuclear magnetism}},}\ }\href {\doibase 10.1016/0029-5582(61)90091-8}
  {\bibfield  {journal} {\bibinfo  {journal} {Am. J. Phys.}\ } (\bibinfo {year}
  {1963}),\ 10.1016/0029-5582(61)90091-8}\BibitemShut {NoStop}%
\bibitem [{\citenamefont {Goold}\ \emph {et~al.}(2016)\citenamefont {Goold},
  \citenamefont {Huber}, \citenamefont {Riera}, \citenamefont {del Rio},\ and\
  \citenamefont {Skrzypczyk}}]{1751-8121-49-14-143001}%
  \BibitemOpen
  \bibfield  {author} {\bibinfo {author} {\bibfnamefont {John}\ \bibnamefont
  {Goold}}, \bibinfo {author} {\bibfnamefont {Marcus}\ \bibnamefont {Huber}},
  \bibinfo {author} {\bibfnamefont {Arnau}\ \bibnamefont {Riera}}, \bibinfo
  {author} {\bibfnamefont {Lídia}\ \bibnamefont {del Rio}}, \ and\ \bibinfo
  {author} {\bibfnamefont {Paul}\ \bibnamefont {Skrzypczyk}},\ }\bibfield
  {title} {\enquote {\bibinfo {title} {The role of quantum information in
  thermodynamics—a topical review},}\ }\href
  {http://stacks.iop.org/1751-8121/49/i=14/a=143001} {\bibfield  {journal}
  {\bibinfo  {journal} {Journal of Physics A: Mathematical and Theoretical}\
  }\textbf {\bibinfo {volume} {49}},\ \bibinfo {pages} {143001} (\bibinfo
  {year} {2016})}\BibitemShut {NoStop}%
\bibitem [{\citenamefont {Batalh\~ao}\ \emph {et~al.}(2014)\citenamefont
  {Batalh\~ao}, \citenamefont {Souza}, \citenamefont {Mazzola}, \citenamefont
  {Auccaise}, \citenamefont {Sarthour}, \citenamefont {Oliveira}, \citenamefont
  {Goold}, \citenamefont {De~Chiara}, \citenamefont {Paternostro},\ and\
  \citenamefont {Serra}}]{PhysRevLett.113.140601}%
  \BibitemOpen
  \bibfield  {author} {\bibinfo {author} {\bibfnamefont {Tiago~B.}\
  \bibnamefont {Batalh\~ao}}, \bibinfo {author} {\bibfnamefont {Alexandre~M.}\
  \bibnamefont {Souza}}, \bibinfo {author} {\bibfnamefont {Laura}\ \bibnamefont
  {Mazzola}}, \bibinfo {author} {\bibfnamefont {Ruben}\ \bibnamefont
  {Auccaise}}, \bibinfo {author} {\bibfnamefont {Roberto~S.}\ \bibnamefont
  {Sarthour}}, \bibinfo {author} {\bibfnamefont {Ivan~S.}\ \bibnamefont
  {Oliveira}}, \bibinfo {author} {\bibfnamefont {John}\ \bibnamefont {Goold}},
  \bibinfo {author} {\bibfnamefont {Gabriele}\ \bibnamefont {De~Chiara}},
  \bibinfo {author} {\bibfnamefont {Mauro}\ \bibnamefont {Paternostro}}, \ and\
  \bibinfo {author} {\bibfnamefont {Roberto~M.}\ \bibnamefont {Serra}},\
  }\bibfield  {title} {\enquote {\bibinfo {title} {Experimental reconstruction
  of work distribution and study of fluctuation relations in a closed quantum
  system},}\ }\href {\doibase 10.1103/PhysRevLett.113.140601} {\bibfield
  {journal} {\bibinfo  {journal} {Phys. Rev. Lett.}\ }\textbf {\bibinfo
  {volume} {113}},\ \bibinfo {pages} {140601} (\bibinfo {year}
  {2014})}\BibitemShut {NoStop}%
\bibitem [{\citenamefont {Hou}\ \emph {et~al.}(2014)\citenamefont {Hou},
  \citenamefont {Sheng}, \citenamefont {Feng},\ and\ \citenamefont
  {Long}}]{Hou2014}%
  \BibitemOpen
  \bibfield  {author} {\bibinfo {author} {\bibfnamefont {Shi~Yao}\ \bibnamefont
  {Hou}}, \bibinfo {author} {\bibfnamefont {Yu~Bo}\ \bibnamefont {Sheng}},
  \bibinfo {author} {\bibfnamefont {Guan~Ru}\ \bibnamefont {Feng}}, \ and\
  \bibinfo {author} {\bibfnamefont {Gui~Lu}\ \bibnamefont {Long}},\ }\bibfield
  {title} {\enquote {\bibinfo {title} {{Experimental optimal single qubit
  purification in an NMR quantum information processor}},}\ }\href {\doibase
  10.1038/srep06857} {\bibfield  {journal} {\bibinfo  {journal} {Sci. Rep.}\
  }\textbf {\bibinfo {volume} {4}} (\bibinfo {year} {2014}),\
  10.1038/srep06857}\BibitemShut {NoStop}%
\bibitem [{\citenamefont {Li}\ \emph {et~al.}(2011)\citenamefont {Li},
  \citenamefont {Yung}, \citenamefont {Chen}, \citenamefont {Lu}, \citenamefont
  {Whitfield}, \citenamefont {Peng}, \citenamefont {Aspuru-Guzik},\ and\
  \citenamefont {Du}}]{Li2011}%
  \BibitemOpen
  \bibfield  {author} {\bibinfo {author} {\bibfnamefont {Zhaokai}\ \bibnamefont
  {Li}}, \bibinfo {author} {\bibfnamefont {Man~Hong}\ \bibnamefont {Yung}},
  \bibinfo {author} {\bibfnamefont {Hongwei}\ \bibnamefont {Chen}}, \bibinfo
  {author} {\bibfnamefont {Dawei}\ \bibnamefont {Lu}}, \bibinfo {author}
  {\bibfnamefont {James~D.}\ \bibnamefont {Whitfield}}, \bibinfo {author}
  {\bibfnamefont {Xinhua}\ \bibnamefont {Peng}}, \bibinfo {author}
  {\bibfnamefont {Al{\'{a}}n}\ \bibnamefont {Aspuru-Guzik}}, \ and\ \bibinfo
  {author} {\bibfnamefont {Jiangfeng}\ \bibnamefont {Du}},\ }\bibfield  {title}
  {\enquote {\bibinfo {title} {{Solving quantum ground-state problems with
  nuclear magnetic resonance}},}\ }\href {\doibase 10.1038/srep00088}
  {\bibfield  {journal} {\bibinfo  {journal} {Sci. Rep.}\ }\textbf {\bibinfo
  {volume} {1}} (\bibinfo {year} {2011}),\ 10.1038/srep00088}\BibitemShut
  {NoStop}%
\bibitem [{\citenamefont {Henry}\ \emph {et~al.}(2006)\citenamefont {Henry},
  \citenamefont {Emerson}, \citenamefont {Martinez},\ and\ \citenamefont
  {Cory}}]{PhysRevA.74.062317}%
  \BibitemOpen
  \bibfield  {author} {\bibinfo {author} {\bibfnamefont {Michael~K.}\
  \bibnamefont {Henry}}, \bibinfo {author} {\bibfnamefont {Joseph}\
  \bibnamefont {Emerson}}, \bibinfo {author} {\bibfnamefont {Rudy}\
  \bibnamefont {Martinez}}, \ and\ \bibinfo {author} {\bibfnamefont {David~G.}\
  \bibnamefont {Cory}},\ }\bibfield  {title} {\enquote {\bibinfo {title}
  {Localization in the quantum sawtooth map emulated on a quantum-information
  processor},}\ }\href {\doibase 10.1103/PhysRevA.74.062317} {\bibfield
  {journal} {\bibinfo  {journal} {Phys. Rev. A}\ }\textbf {\bibinfo {volume}
  {74}},\ \bibinfo {pages} {062317} (\bibinfo {year} {2006})}\BibitemShut
  {NoStop}%
\bibitem [{\citenamefont {Cummins}\ \emph {et~al.}(2002)\citenamefont
  {Cummins}, \citenamefont {Jones}, \citenamefont {Furze}, \citenamefont
  {Soffe}, \citenamefont {Mosca}, \citenamefont {Peach},\ and\ \citenamefont
  {Jones}}]{PhysRevLett.88.187901}%
  \BibitemOpen
  \bibfield  {author} {\bibinfo {author} {\bibfnamefont {Holly~K.}\
  \bibnamefont {Cummins}}, \bibinfo {author} {\bibfnamefont {Claire}\
  \bibnamefont {Jones}}, \bibinfo {author} {\bibfnamefont {Alistair}\
  \bibnamefont {Furze}}, \bibinfo {author} {\bibfnamefont {Nicholas~F.}\
  \bibnamefont {Soffe}}, \bibinfo {author} {\bibfnamefont {Michele}\
  \bibnamefont {Mosca}}, \bibinfo {author} {\bibfnamefont {Josephine~M.}\
  \bibnamefont {Peach}}, \ and\ \bibinfo {author} {\bibfnamefont {Jonathan~A.}\
  \bibnamefont {Jones}},\ }\bibfield  {title} {\enquote {\bibinfo {title}
  {Approximate quantum cloning with nuclear magnetic resonance},}\ }\href
  {\doibase 10.1103/PhysRevLett.88.187901} {\bibfield  {journal} {\bibinfo
  {journal} {Phys. Rev. Lett.}\ }\textbf {\bibinfo {volume} {88}},\ \bibinfo
  {pages} {187901} (\bibinfo {year} {2002})}\BibitemShut {NoStop}%
\bibitem [{\citenamefont {Kosloff}(2013)}]{e15062100}%
  \BibitemOpen
  \bibfield  {author} {\bibinfo {author} {\bibfnamefont {Ronnie}\ \bibnamefont
  {Kosloff}},\ }\bibfield  {title} {\enquote {\bibinfo {title} {Quantum
  thermodynamics: A dynamical viewpoint},}\ }\href {\doibase 10.3390/e15062100}
  {\bibfield  {journal} {\bibinfo  {journal} {Entropy}\ }\textbf {\bibinfo
  {volume} {15}},\ \bibinfo {pages} {2100--2128} (\bibinfo {year}
  {2013})}\BibitemShut {NoStop}%
\bibitem [{\citenamefont {Vinjanampathy}\ and\ \citenamefont
  {Anders}(2016)}]{doi:10.1080/00107514.2016.1201896}%
  \BibitemOpen
  \bibfield  {author} {\bibinfo {author} {\bibfnamefont {Sai}\ \bibnamefont
  {Vinjanampathy}}\ and\ \bibinfo {author} {\bibfnamefont {Janet}\ \bibnamefont
  {Anders}},\ }\bibfield  {title} {\enquote {\bibinfo {title} {Quantum
  thermodynamics},}\ }\href {\doibase 10.1080/00107514.2016.1201896} {\bibfield
   {journal} {\bibinfo  {journal} {Contemporary Physics}\ }\textbf {\bibinfo
  {volume} {57}},\ \bibinfo {pages} {545--579} (\bibinfo {year}
  {2016})}\BibitemShut {NoStop}%
\bibitem [{\citenamefont {Redfield}(1957)}]{5392713}%
  \BibitemOpen
  \bibfield  {author} {\bibinfo {author} {\bibfnamefont {A.~G.}\ \bibnamefont
  {Redfield}},\ }\bibfield  {title} {\enquote {\bibinfo {title} {On the theory
  of relaxation processes},}\ }\href {\doibase 10.1147/rd.11.0019} {\bibfield
  {journal} {\bibinfo  {journal} {IBM Journal of Research and Development}\
  }\textbf {\bibinfo {volume} {1}},\ \bibinfo {pages} {19--31} (\bibinfo {year}
  {1957})}\BibitemShut {NoStop}%
\end{thebibliography}%


\end{document}